\documentclass[preprint,pre,floats,aps,amsmath,amssymb, standalone]{revtex4}
\usepackage{braket}
\usepackage{graphicx}
\usepackage{bm}

\begin{document}

\title{Study of spin-spin correlations between quark and a spin-$\frac{1}{2}$ composite system}
\author{Satvir Kaur and Harleen Dahiya}
\affiliation{Department of Physics,\\ Dr. B.R. Ambedkar National
	Institute of Technology,\\ Jalandhar, 144011, India}
%\date{\today}

\begin{abstract}
We study the correlation between the fermion composite system and quark spins by using the light-cone quark-diquark model. We do the calculations for $u$-quark and $d$-quark in the fermion system by considering different polarization configrations of both. The contribution from scalar and axial-vector diquarks is taken into account. The overlap representation of light-front wavefunctions is used for the calculations. The spin-spin correlations for $u$ and $d$ quarks are presented in transverse impact-parameter plane and transverse momentum plane as well.
\end{abstract}

\maketitle

\section{Introduction}
To get  precise information of hadrons in terms of its constituents, Wigner distributions of quark and gluon were introduced by Ji \cite{ji, ji1}. Wigner distribution is a quantum phase-space distribution concealing the joint position and momentum space distribution on the internal structure of the hadron. As these distributions are quasi-probabilistic distributions, one cannot measure them directly. Applying certain limits on Wigner distributions provide the probabilistic three-dimensional distributions namely generalized parton distributions (GPDs) \cite{diehl, xji, belitsky, garcon} and transverse momentum-dependent distributions (TMDs) \cite{tmd-bachhetta, tmd-meissner, tmd-xiao, collins}. The reduction to GPDs is based on the integration of five-dimensional Wigner distributions over transverse momentum at zero skewness. While at the forward limit, i.e. when there is no momentum transfer from intial to final state of hadron $({\bf \Delta}_\perp=0)$, the TMDs can be obtained by integrating Wigner distributions over transverse impact-parameter co-ordinates. Further integrating GPDs upon certain limits leads to obtaining parton distribution functions (PDFs), charge distributions, form factors etc. \cite{ff, ff1, ff2, ff3, ff4, ff5, ff6, ff7}. Wigner distributions are also supportive for evaluating the spin-spin correlations between a spin-$\frac{1}{2}$ composite system and a quark inside the fermion system. 
%The accomplishment of Wigner distribution in other fields like quantum molecular dynamics, non-linear dynamics, signal analysis, heavy ion collision etc. \cite{signal, signal1, signal2} has been achieved. 
Theoretical studies on quark and gluon Wigner distributions in spin-$\frac{1}{2}$ and spin-$0$ composite systems have been successfully carried out in Ref. \cite{wd, wd1, wd2, wd3, wd4, wd5, wd6, wd7, wd8}.

The spectator model formulated in light-cone framework \cite{lc_dirac, lc_harindranath, lc_brodsky, lc_zhang} is used to evaluate the Wigner distributions as it is successful in evaluating T-even and T-odd TMDs of the proton \cite{tmd-bachhetta}. The model is successful in explaining the standard parton distribution functions, quasi-parton distribution functions \cite{pdf-spectator-model}. Further, in Ref. \cite{gpd-spectator-model}, the authors analyse the agreement of quasi-GPDs with the standard GPDs. The concept of quasi-PDF is carried out in Ref. \cite{qpdf, qpdf1, qpdf2, qpdf3, qpdf4, qpdf5}. Since, the fermion composite system is considered to be a bound state of three quarks i.e. $uud$, the spectators are assumed to be scalar or axial-vector depending upon the spin i.e. either spin-$0$ or spin-$1$. In this work, we investigate the correlation between the quark spin and spin of fermion system by using the Wigner distributions evaluated in Ref. \cite{wd8}. The quark Wigner distributions were calculated by considering  different configuration combinations of quark spin direction and proton spin direction. The overlap representation of light-front wavefunctions is taken into account to evaluate the Wigner operators having different cases, depending upon the polarization of quark i.e. either unpolarized, longitudinally-polarized or transversely-polarized. The Wigner operator is associated with the Wigner distributions by a Fourier transformation of total momentum transferred to the final state of the system. Furthermore, we include the longitudinal polarization vector into the LFWFs along with the transverse polarization vector, and evaluate the Wigner distribution using the overlap form of these LFWFs.  Afterwards, we take the difference between both respective Wigner distributions i.e. including the longitudinal polarization vector $(\rho^{(a)l})$ and by not including the longitudinal polarization vector $(\rho^{(a)})$. %we calculate the difference between Wigner distributions for axial-vector diquark evaluated by using the LFWFs by considering longitudinal polarization vector $(W^{(a)l})$ and by not considering longitudinal polarization vector $(W^{(a)})$ into account i.e. $(W^{(a)l}-W^{(a)})$.
 The aim behind the determination of difference between the Wigner distributions is just to get the effect of longitudinal polarization vector in LFWFs.  

The plan of the paper is as follows. In Section-II, we briefly discuss about the light-front quark-diquark model used. In Section III, the definitions of Wigner distribution in terms of polarization configurations of quark and spin-$\frac{1}{2}$ composite system are given. We also introduce various spin-spin correlations between the quark and composite system in this section. Further, in Section IV, we evaluate the difference between the correlators related to the Wigner distributions in terms of overlap form of LFWFs for the cases where longitudinal polarization vector is taken into account as well as the case without longitudinal polarization vector. Also, the results of different spin-spin correlations are discussed. At last, the summary and conclusions are presented in Section V.

\section{Light-front quark-diquark model}
In the light-front quark-diquark model, the spin-$\frac{1}{2}$ composite system is considered to be a bound state of a quark and a diquark. In this model, a valence quark interacts with the external photon, and the other two valence quarks bound together are treated as a single diquark state. Here, the diquark can be scalar (spin-$0$) or axial-vector (spin-$1$). The composite spin-$\frac{1}{2}$ particle state $\ket{\Psi;S}$ is defined as
\begin{eqnarray}
\ket{\Psi;\pm}=c_s\ket{u \ s^0}^{\pm}+c_a\ket{u \ a^0}^{\pm}+c_a'\ket{d \ a^1}^{\pm}.
\end{eqnarray} 
Here, the scalar-isoscalar diquark state, vector-isoscalar diquark state and vector-isovector diquark state are denoted by $\ket{u \ s}^0$, $\ket{u \ a^0}$ and $\ket{d \ a^1}$ respectively.

The hadronic light-cone Fock state $|\Psi (P^+, \mathbf{P}_\perp, S_z) \rangle$ expansion in terms of constituent eigenstates, is defined as  \cite{brodsky}
\begin{eqnarray}
|\Psi(P^+, \mathbf{P}_\perp, S_z) \rangle
   &=&\sum_{n,\lambda_i}\int\prod_{i=1}^n \frac{\mathrm{d} x_i \mathrm{d}^2
        \mathbf{p}_{\perp i}}{\sqrt{x_i}~16\pi^3}
        16\pi^3 \delta\Big(1-\sum_{i=1}^n x_i\Big)\delta^{(2)}\Big(\sum_{i=1}^n \mathbf{p}_{\perp i}\Big) \nonumber\\
       &\times& | n;x_i P^+, x_i \mathbf{P}_\perp+\mathbf{p}_{\perp i},
        \lambda_i \rangle
        \psi_{n/M}(x_i,\mathbf{p}_{\perp i},\lambda_i), \nonumber\\
        \label{hadron_eqn}
\end{eqnarray}
where $x_i=\frac{p_i^+}{P^+}$ is light-cone momentum fraction and ${\bf p}_{\perp i}$ is the relative momentum of the $i$th constituent of the hadron. The helicity of $ith$ constituent is denoted by $\lambda_i$. 
The Fock states of $n$-particle are normalized as follows
\begin{eqnarray}
\langle{n;{p'}_i^+, {\textbf{p}'}_{\perp i},\lambda'_i} | {n;{p}_i^+, {\textbf{p}}_{\perp i},\lambda_i}\rangle &=&\prod_{i=1}^n 16 \pi^3 p_i^+\delta({p'}_i^+ -p_i^+)
\delta^{(2)}({\textbf{p}'}_{\perp i}-{\textbf{p}}_{\perp i})\delta_{\lambda'_i \lambda_i}.
\end{eqnarray}

As the system is considered as a two-particle system (a quark and a diquark), therefore, by substituting $n=2$, the Fock state expansion for scalar diquark $(\ket{u \ s^0})$ leads to 
\begin{eqnarray}
\ket{u \ s^0 (P^+, {\bf p}_\perp)}^{\pm}=\sum_{\lambda_q} \int \frac{dx d^2{\bf p}_\perp}{\sqrt{x(1-x)} 16\pi^3} \psi^\pm_ {\lambda_q}(x, {\bf p}_\perp)\ket{x P^+, {\bf p}_\perp,\lambda_q}.
\label{scalar-diquark}
\end{eqnarray}  
Similarly, the expansion of axial-vector diquark component is expressed as
\begin{eqnarray}
\ket{\mu \ V (P^+, {\bf p}_\perp)}^{\pm}=\sum_{\lambda_q, \lambda_D} \int \frac{dx d^2{\bf p}_\perp}{\sqrt{x(1-x)} 16\pi^3} \psi^\pm_ {\lambda_q \lambda_D}(x, {\bf p}_\perp)\ket{x P^+, {\bf p}_\perp,\lambda_q, \lambda_D},
\label{axial-vector-diquark}
\end{eqnarray}
where the respective helicities of quark and diquark are denoted by $\lambda_q$ and $\lambda_D$. Here, $\mu$ can be $u$-quark or $d$-quark and $V$ denotes the axial vector diquark, either isoscalar or isovector.

The wavefunctions related to the scalar diquark are defined as \cite{tmd-bachhetta}
\begin{eqnarray}
\psi^{+}_{+} (x,\textbf{p}_{\perp})&=&\frac{m+ x M}{x}\,\varphi(x,\textbf{p}_{\perp}) ,  \nonumber\\
\psi^{+}_{-} (x,\textbf{p}_{\perp}) &=& -\frac{p_x+i p_y}{x}\,\varphi(x,\textbf{p}_{\perp}), \nonumber\\
\psi^{-}_{+} (x,\textbf{p}_{\perp}) &=& \frac{p_x-p_y}{x} \ \varphi(x,\textbf{p}_{\perp}),\nonumber \\
\psi^{-}_{-} (x,\textbf{p}_{\perp}) &=& \frac{m+xM}{x}\ \varphi(x,\textbf{p}_{\perp}),
\end{eqnarray}
with
\begin{eqnarray}
\varphi(x,\textbf{p}_{\perp}) &=& -\frac{g_s}{\sqrt{1-x}}\, 
    \frac{x (1-x)}{\textbf{p}_{\perp}^2+[xM^{2}_{s}+(1-x)m^{2}-x(1-x)M^{2}]} \;.
    \label{scalar:var}
\label{eq:lcwf-s2}
\end{eqnarray}
Similarly, the wavefunctions related to axial-vector diquark are defined as
\begin{eqnarray}
\psi^{+}_{+\frac{1}{2}+1}(x,{\bf p_\perp})&=& \frac{(p_{x}-i p_y)}{x(1-x)}\phi(x,\textbf{p}_{\perp}),\nonumber\\
\psi^{+}_{+\frac{1}{2}-1}(x,{\bf p_\perp})&=& -\frac{(p_x+i p_y)}{(1-x)}\phi(x,\textbf{p}_{\perp}),\nonumber\\
\psi^{+}_{-\frac{1}{2}+1}(x,{\bf p_\perp})&=& \frac{(m+xM)}{x} \phi(x,\textbf{p}_{\perp}),\nonumber\\
\psi^{+}_{-\frac{1}{2}-1}(x,{\bf p_\perp})&=& 0,\\
\psi^{-}_{+\frac{1}{2}+1}(x,{\bf p_\perp})&=& 0,\nonumber\\
\psi^{-}_{+\frac{1}{2}-1}(x,{\bf p_\perp})&=& -\frac{(m+xM)}{x} \phi(x,\textbf{p}_{\perp}) ,\nonumber\\
\psi^{-}_{-\frac{1}{2}+1}(x,{\bf p_\perp})&=& -\frac{(p_x-i p_y)}{(1-x)}\phi(x,\textbf{p}_{\perp}),\nonumber\\
\psi^{-}_{-\frac{1}{2}-1}(x,{\bf p_\perp})&=& \frac{(p_{x}+i p_y)}{x(1-x)}\phi(x,\textbf{p}_{\perp}).
\label{axial-vector-wavefunctions}
\end{eqnarray}
\begin{eqnarray}
\phi(x,\textbf{p}_{\perp})=-\frac{g_{a}}{\sqrt{1-x}}\frac{x(1-x)}{\textbf{p}^{2}_{\perp}+[xM^{2}_{a}+(1-x)m^{2}-x(1-x)M^{2}]}. \label{phi}
\end{eqnarray}

The above wavefunctions for axial-vector diquark are defined corresponding to the light-cone transverse polarization vectors satisfying $\epsilon(\pm) . \epsilon^{*}(\pm)=-1$, $\epsilon(\pm) . \epsilon^{*}(\mp)=0$ and $(P-p). \epsilon(\pm)=0$ given below :
\begin{eqnarray}
\epsilon(P-p,+)=\bigg[\frac{p_x+i p_y}{\sqrt{2}(1-x)P^+},0,-\frac{1}{\sqrt{2}},-\frac{i}{\sqrt{2}}\bigg],\\
\epsilon(P-p,-)=\bigg[-\frac{p_x-i p_y}{\sqrt{2}(1-x)P^+},0,\frac{1}{\sqrt{2}},-\frac{i}{\sqrt{2}}\bigg].
\end{eqnarray}
In addition to this, the third longitudinal polarization vector is also included with the transverse polarization vectors. It satisfies $\epsilon(0).\epsilon^{*}(0)=-1, \epsilon(0). \epsilon^*(\pm)=0$, and $(P-p).\epsilon(0)=0$.
\begin{eqnarray}
\epsilon(P-p,0)=\frac{1}{M_a}\bigg[\frac{{\bf p}^2_\perp-M^2_a}{2(1-x)P^+},(1-x)P^+,-p_x,-p_y\bigg].
\end{eqnarray}

The light-cone wavefunctions corresponding to above longitudinal polarization vector are defined as
\begin{eqnarray}
\psi^+_{+0}(x,{\bf p}_\perp)&=&\frac{{\bf p}^2_\perp-xM_a^2-mM(1-x)^2}{\sqrt{2}x(1-x)M_a}\phi(x,\textbf{p}_{\perp}),\nonumber\\
\psi^+_{-0}(x,{\bf p}_\perp)&=&\frac{(m+M)}{\sqrt{2}M_a}(p_x+i p_y)\phi(x,\textbf{p}_{\perp}),\nonumber\\
\psi^-_{+0}(x,{\bf p}_\perp)&=&\frac{(m+M)}{\sqrt{2}M_a}(p_x-i p_y)\phi(x,\textbf{p}_{\perp}),\nonumber\\
\psi^-_{-0}(x,{\bf p}_\perp)&=&-\frac{{\bf p}^2_\perp-xM_a^2-mM(1-x)^2}{\sqrt{2}x(1-x)M_a}\phi(x,\textbf{p}_{\perp}),
\label{longitudinal-polarization-vector}
\end{eqnarray}
where $M_a$, $M$ and $m$ are axial-vector diquark mass, spin-$\frac{1}{2}$ particle mass and constituent quark mass respectively.

\section{Wigner distributions and spin-spin correlations}
The five-dimensional Wigner distribution of quark, also known as quantum phase-space distribution, is defined as \cite{wigner-lorce}
\begin{eqnarray}
\rho^{[\Gamma]}({\bf b}_\perp,{\bf k}_\perp, x,S)\equiv \int \frac{d^2 {\bf \Delta}_\perp}{(2 \pi)^2} e^{-i {\bf \Delta}_\perp . {\bf b}_\perp} {W}^{[\Gamma]}({\bf \Delta}_\perp, {\bf k}_\perp, x, S),\nonumber\\
\label{wigner}
\end{eqnarray}
where the correlator ${W}^{[\Gamma]}(\Delta_\perp, {\bf k_\perp},x;S)$ is
\begin{eqnarray}
&&{W}^{[\Gamma]}(\Delta_\perp, {\bf k_\perp},x;S)=\frac{1}{2} \int \frac{dz^- d^2 z_\perp}{(2 \pi)^3} e^{i k \cdot z}  \bigg\langle{P'';S}\bigg| \bar{\psi}\bigg(-\frac{z}{2}\bigg) \Gamma \mathcal{W}_{[-\frac{z}{2},\frac{z}{2}]}\psi\bigg(\frac{z}{2}\bigg)\bigg|{P';S}\bigg\rangle\Bigg\vert_{z^{+}=0}.\nonumber\\
\label{wigner_correlator}
\end{eqnarray} 
Here, $\Gamma$ defines the Dirac gamma matrices and $\Gamma=\gamma^+,\gamma^+\gamma^5, i\sigma^{j+}\gamma^5$. The state of the composite system are defined in Eqs. (\ref{scalar-diquark}) and (\ref{axial-vector-diquark}) based on whether the diquark is scalar or axial-vector. By substituting Eqs. (\ref{scalar-diquark}) and (\ref{axial-vector-diquark}) in Eq. (\ref{wigner_correlator}), one can get the overlap form of the Wigner distribution. 

The phase-space distributions based on the configurations of various polarizations i.e. $\rho_{XY}$, where $X$ defines the polarization of composite system and $Y$ stands for the polarization of quark, are defined as \cite{wd5, wd6, wd8}
 \begin{eqnarray}
\rho_{UU}({\bf b_\perp},{\bf p_\perp},x)&=& \frac{1}{2} \Big[\rho^{[\gamma^+]}({\bf b_\perp},{\bf p_\perp},x;+\hat{S}_z) + \rho^{[\gamma^+]}({\bf b_\perp},{\bf p_\perp},x;-\hat{S}_z)\Big],\\
\rho_{UL}({\bf b_\perp},{\bf p_\perp},x)&=& \frac{1}{2} \Big[\rho^{[\gamma^+ \gamma_5]}({\bf b_\perp},{\bf p_\perp},x;+\hat{S}_z) + \rho^{[\gamma^+ \gamma_5]}({\bf b_\perp},{\bf p_\perp},x;-\hat{S}_z)\Big],
\label{ul}\\
\rho^{j}_{UT}({\bf b_\perp},{\bf p_\perp},x)&=& \frac{1}{2} \Big[\rho^{[i\sigma^{+j} \gamma_5]} ({\bf b_\perp},{\bf p_\perp},x;+\hat{S}_z) + \rho^{[i\sigma^{+j} \gamma_5]}({\bf b_\perp},{\bf p_\perp},x;-\hat{S}_z)\Big],
\label{ut}\\
\rho_{LU}({\bf b_\perp},{\bf p_\perp},x)&=& \frac{1}{2}\Big[\rho^{[\gamma^+]} ({\bf b_\perp},{\bf p_\perp},x;+\hat{S}_z) - \rho^{[\gamma^+]} ({\bf b_\perp},{\bf p_\perp},x;-\hat{S}_z)\Big],
\label{lu}\\
\rho_{LL}({\bf b_\perp},{\bf p_\perp},x)&=&\frac{1}{2} \Big[\rho^{[\gamma^+ \gamma_5]} ({\bf b_\perp},{\bf p_\perp},x;+\hat{S}_z) - \rho^{[\gamma^+ \gamma_5]} ({\bf b_\perp},{\bf p_\perp},x;-\hat{S}_z)\Big], 
\label{ll}\\
\rho^{j}_{LT}({\bf b_\perp},{\bf p_\perp},x)&=& \frac{1}{2}\Big[\rho^{[i \sigma^{+j} \gamma_5]} ({\bf b_\perp},{\bf p_\perp},x;+\hat{S}_z) - \rho^{[i \sigma^{+j} \gamma_5]} ({\bf b_\perp},{\bf p_\perp},x;-\hat{S}_z)\Big],
\label{lt}\\
\rho^i_{TU}({\bf b_\perp},{\bf p_\perp},x)&=& \frac{1}{2} \Big[\rho^{[\gamma^{+} ]} ({\bf b_\perp},{\bf p_\perp},x;+\hat{S}_i) - \rho^{[\gamma^{+}]} ({\bf b_\perp},{\bf p_\perp},x;-\hat{S}_i)\Big],
\label{tu}\\
\rho^i_{TL}({\bf b_\perp},{\bf p_\perp},x)&=& \frac{1}{2} \Big[\rho^{[\gamma^{+} \gamma_5 ]} ({\bf b_\perp},{\bf p_\perp},x;+\hat{S}_i) - \rho^{[\gamma^{+} \gamma_5]} ({\bf b_\perp},{\bf p_\perp},x;-\hat{S}_i)\Big],
\label{tl}\\
\rho_{TT}({\bf b_\perp},{\bf p_\perp},x)&=& \frac{1}{2} \delta_{ij} \Big[\rho^{[i \sigma^{+j} \gamma_5]}({\bf b_\perp},{\bf p_\perp},x;+\hat{S}_i) - \rho^{[i \sigma^{+j} \gamma_5]}({\bf b_\perp},{\bf p_\perp},x;-\hat{S}_i)\Big],
\label{tt}
\end{eqnarray}
and finally the pretzelous  Wigner distribution as
\begin{eqnarray}
\rho^{\perp}_{TT}({\bf b_\perp},{\bf p_\perp},x)&=& \frac{1}{2} \epsilon_{ij} \Big[\rho^{[i \sigma^{+j} \gamma_5]}({\bf b_\perp},{\bf p_\perp},x;+\hat{S}_i) - \rho^{[i \sigma^{+j} \gamma_5]}({\bf b_\perp},{\bf p_\perp},x;-\hat{S}_i)\Big].
\label{pret}
\end{eqnarray}
Here, in the subscript of Wigner distributions, $U$, $L$ and $T$ explains whether the quark or a fermion composite system is unpolarized, longitudinally-polarized or transversely-polarized.
  
To extract information about the correlation between quark spin and fermion system spin, the Wigner distributions of quarks in the proton having different helicities are evaluated. For $\Gamma=\gamma^+\frac{1+\lambda \gamma^5}{2}$ and $\overrightarrow{S}=\Lambda \hat{S}_z$, the longitudinal Wigner distribution of the quark in the fermion system having helicities $\lambda$ and $\Lambda$ respectively, is defined as 
\begin{eqnarray}
\rho_{\Lambda \lambda}({\bf b}_\perp,{\bf p}_\perp, x)&=&\frac{1}{2}[\rho^{[\gamma^+]}({\bf b}_\perp, {\bf p}_\perp, x, \Lambda \hat{S}_z)+\lambda\rho^{[\gamma^+ \gamma^5]}({\bf b}_\perp, {\bf p}_\perp, x, \Lambda \hat{S}_z)].
\end{eqnarray}
 The above equation can be expressed in terms of polarization configurations of quark and proton as 
\begin{eqnarray}
\rho_{\Lambda \lambda}({\bf b}_\perp,{\bf p}_\perp, x)&=&\frac{1}{2}[\rho_{UU}({\bf b}_\perp,{\bf p}_\perp, x)+\Lambda \rho_{LU}({\bf b}_\perp,{\bf p}_\perp, x)+\lambda \rho_{UL}({\bf b}_\perp,{\bf p}_\perp, x)\nonumber\\
&&+\Lambda \lambda \rho_{LL}({\bf b}_\perp,{\bf p}_\perp, x)].
\label{longitudinal-wigner}
\end{eqnarray}
For the quark Wigner distributions, considering the spin directions of quark and composite system to be in the longitudinal direction, the helicities $\Lambda$ and $\lambda$ take different forms i.e. $\Lambda=\uparrow, \downarrow$ and $\lambda=\uparrow, \downarrow$. 

Similar to the longitudinal Wigner distributions, the Wigner distributions for quark having the transverse polarization $\lambda_\perp=\Uparrow,\Downarrow $ in the composite system having transverse polarization $\Lambda_\perp=\Uparrow, \Downarrow$, for $\Gamma=\frac{\gamma^+ +\Lambda_\perp{i \sigma^{j+}\gamma^5}}{2}$ and $\overrightarrow{S}=\Lambda_\perp \hat{S}_i$ is given as
\begin{eqnarray}
\rho_{\Lambda_\perp \lambda_\perp}({\bf b}_\perp,{\bf p}_\perp, x)&=&\frac{1}{2}[\rho^{[\gamma^+]}({\bf b}_\perp, {\bf p}_\perp, x, \Lambda_\perp \hat{S}_i)+\Lambda_\perp\rho^{[i\sigma^{j+}\gamma^+]}({\bf b}_\perp, {\bf p}_\perp, x, \Lambda_\perp \hat{S}_i)].
\label{transverse-wigner}
\end{eqnarray}
In terms of polarization configrations, the above equation can be expressed as
\begin{eqnarray}
\rho^i_{\Lambda_\perp \lambda_\perp}({\bf b}_\perp,{\bf p}_\perp, x)&=&\frac{1}{2}[\rho_{UU}({\bf b}_\perp,{\bf p}_\perp, x)+\Lambda_\perp \rho^i_{TU}({\bf b}_\perp,{\bf p}_\perp, x)+\lambda_\perp \rho^i_{UT}({\bf b}_\perp,{\bf p}_\perp, x)\nonumber\\
&&+\Lambda_\perp \lambda_\perp \rho_{TT}({\bf b}_\perp,{\bf p}_\perp, x)].
\end{eqnarray}

Further, for the quark having spin in longitudinal direction and fermion system spin in transverse direction and vice-versa, the respective Wigner distributions $\rho^i_{\Lambda_\perp \lambda}$ and $\rho^j_{\Lambda \lambda_\perp}$ are defined as 
\begin{eqnarray}
\rho^i_{\Lambda_\perp \lambda}({\bf b}_\perp,{\bf p}_\perp, x)&=&\frac{1}{2}[\rho_{UU}({\bf b}_\perp,{\bf p}_\perp, x)+\Lambda_\perp \rho^i_{TU}({\bf b}_\perp,{\bf p}_\perp, x)+\lambda \rho_{UL}({\bf b}_\perp,{\bf p}_\perp, x)\nonumber\\
&&+\Lambda_\perp \lambda \rho^i_{TL}({\bf b}_\perp,{\bf p}_\perp, x)],
\label{trans-longi-distribution}
\end{eqnarray}
and
\begin{eqnarray}
\rho^j_{\Lambda \lambda_\perp}({\bf b}_\perp,{\bf p}_\perp, x)&=&\frac{1}{2}[\rho_{UU}({\bf b}_\perp,{\bf p}_\perp, x)+\Lambda \rho_{LU}({\bf b}_\perp,{\bf p}_\perp, x)+\lambda_\perp \rho^j_{UT}({\bf b}_\perp,{\bf p}_\perp, x)\nonumber\\
&&+\Lambda \lambda_\perp \rho^j_{LT}({\bf b}_\perp,{\bf p}_\perp, x)].
\label{longi-trans-distribution}
\end{eqnarray}

\section{Results}
Using the overlap form of LFWFs for axial-vector diquark, the difference between the Wigner operators for the case where longitudinal polarization vector is included and for the case where the longitudinal polarization vector is not included (from Eqs. (\ref{axial-vector-wavefunctions}) and (\ref{longitudinal-polarization-vector})), we have
\begin{eqnarray}
W^{(a)l}_{UU}-W^{(a)}_{UU}&=&\frac{1}{16\pi^3}\bigg[\frac{({\bf p}''^2_\perp-x M^2_a-x M(1-x)^2)({\bf p}'^2_\perp-x M^2_a-x M(1-x)^2)}{x^2(1-x)^2}\nonumber\\
&+&\frac{(m+M)}{2 M^2_a x^2}\bigg({\bf p}^2_\perp-\frac{(1-x)^2}{4}{\bf \Delta}^2_\perp \bigg)\bigg] \phi^\dagger(x,{\bf p}''_\perp)\phi(x,{\bf p}'_\perp),\\
W^{(a)l}_{UL}-W^{(a)}_{UL}&=& \frac{i}{16 \pi^3}\frac{(m+M)^2}{2 M_a^2 x^2} (1-x)(p_x \Delta_y-p_y \Delta_x)\phi^\dagger(x,{\bf p}''_\perp)\phi(x,{\bf p}'_\perp),\\
W^{(a)l}_{UT}-W^{(a)}_{UT}&=& \frac{i}{16 \pi^3} \frac{(m+M)}{2 M^2_a x^2}\bigg[({\bf p}''^2_\perp-x M^2_a-x M(1-x)^2)\bigg(p_y-\frac{(1-x)}{2}\Delta_y\bigg)\nonumber\\
&+&({\bf p}'^2_\perp-x M^2_a-x M(1-x)^2)\bigg(p_y+\frac{(1-x)}{2}\Delta_y\bigg)\bigg]\phi^\dagger(x,{\bf p}''_\perp)\phi(x,{\bf p}'_\perp),\\
W^{(a)l}_{LU}-W^{(a)}_{LU}&=&-\frac{i}{16 \pi^3}\frac{(m+M)^2}{2 M_a^2 x^2} (1-x)(p_x \Delta_y-p_y \Delta_x)\phi^\dagger(x,{\bf p}''_\perp)\phi(x,{\bf p}'_\perp),\\
W^{(a)l}_{LL}-W^{(a)}_{LL}&=&\frac{1}{16\pi^3}\bigg[\frac{({\bf p}''^2_\perp-x M^2_a-x M(1-x)^2)({\bf p}'^2_\perp-x M^2_a-x M(1-x)^2)}{x^2(1-x)^2}\nonumber\\
&-&\frac{(m+M)}{2 M^2_a x^2}\bigg({\bf p}^2_\perp-\frac{(1-x)^2}{4}{\bf \Delta}^2_\perp \bigg)\bigg] \phi^\dagger(x,{\bf p}''_\perp)\phi(x,{\bf p}'_\perp),\\
W^{(a)l}_{LT}-W^{(a)}_{LT}&=&\frac{1}{16\pi^3}\frac{(m+M)}{2 M_a^2 x^2}\bigg[({\bf p}''^2_\perp-x M^2_a-x M(1-x)^2)\bigg(p_x-\frac{(1-x)}{2}\Delta_x\bigg)\nonumber\\
&+&({\bf p}'^2_\perp-x M^2_a-x M(1-x)^2)\bigg(p_x+\frac{(1-x)}{2}\Delta_x\bigg)\bigg]\phi^\dagger(x,{\bf p}''_\perp)\phi(x,{\bf p}'_\perp),\\
W^{(a)l}_{TU}-W^{(a)}_{TU}&=&\frac{1}{16\pi^3}\frac{(m+M)}{2 M_a^2 x^2}\bigg[({\bf p}''^2_\perp-x M^2_a-x M(1-x)^2)\bigg(p_x-\frac{(1-x)}{2}\Delta_x\bigg)\nonumber\\
&-&({\bf p}'^2_\perp-x M^2_a-x M(1-x)^2)\bigg(p_x+\frac{(1-x)}{2}\Delta_x\bigg)\bigg]\phi^\dagger(x,{\bf p}''_\perp)\phi(x,{\bf p}'_\perp),\\
W^{(a)l}_{TL}-W^{(a)}_{TL}&=& -\frac{i}{16 \pi^3} \frac{(m+M)}{2 M^2_a x^2}\bigg[({\bf p}''^2_\perp-x M^2_a-x M(1-x)^2)\bigg(p_y-\frac{(1-x)}{2}\Delta_y\bigg)\nonumber\\
&+&({\bf p}'^2_\perp-x M^2_a-x M(1-x)^2)\bigg(p_y+\frac{(1-x)}{2}\Delta_y\bigg)\bigg]\phi^\dagger(x,{\bf p}''_\perp)\phi(x,{\bf p}'_\perp),\\
W^{(a)l}_{TT}-W^{(a)}_{TT}&=& -\frac{1}{16 \pi^3}\bigg[\frac{({\bf p}''^2_\perp-x M^2_a-x M(1-x)^2)({\bf p}'^2_\perp-x M^2_a-x M(1-x)^2)}{x^2(1-x)^2}\nonumber\\
&+&\frac{(m+x M)^2}{2 M_a^2 x^2}\bigg((p_x^2-p_y^2)-\frac{(1-x)^2}{4}(\Delta_x^2-\Delta_y^2)\bigg)\bigg]\phi^\dagger(x,{\bf p}''_\perp)\phi(x,{\bf p}'_\perp).
\end{eqnarray}
\begin{figure}
\centering
\begin{minipage}[c]{1\textwidth}
(a)\includegraphics[width=.4\textwidth]{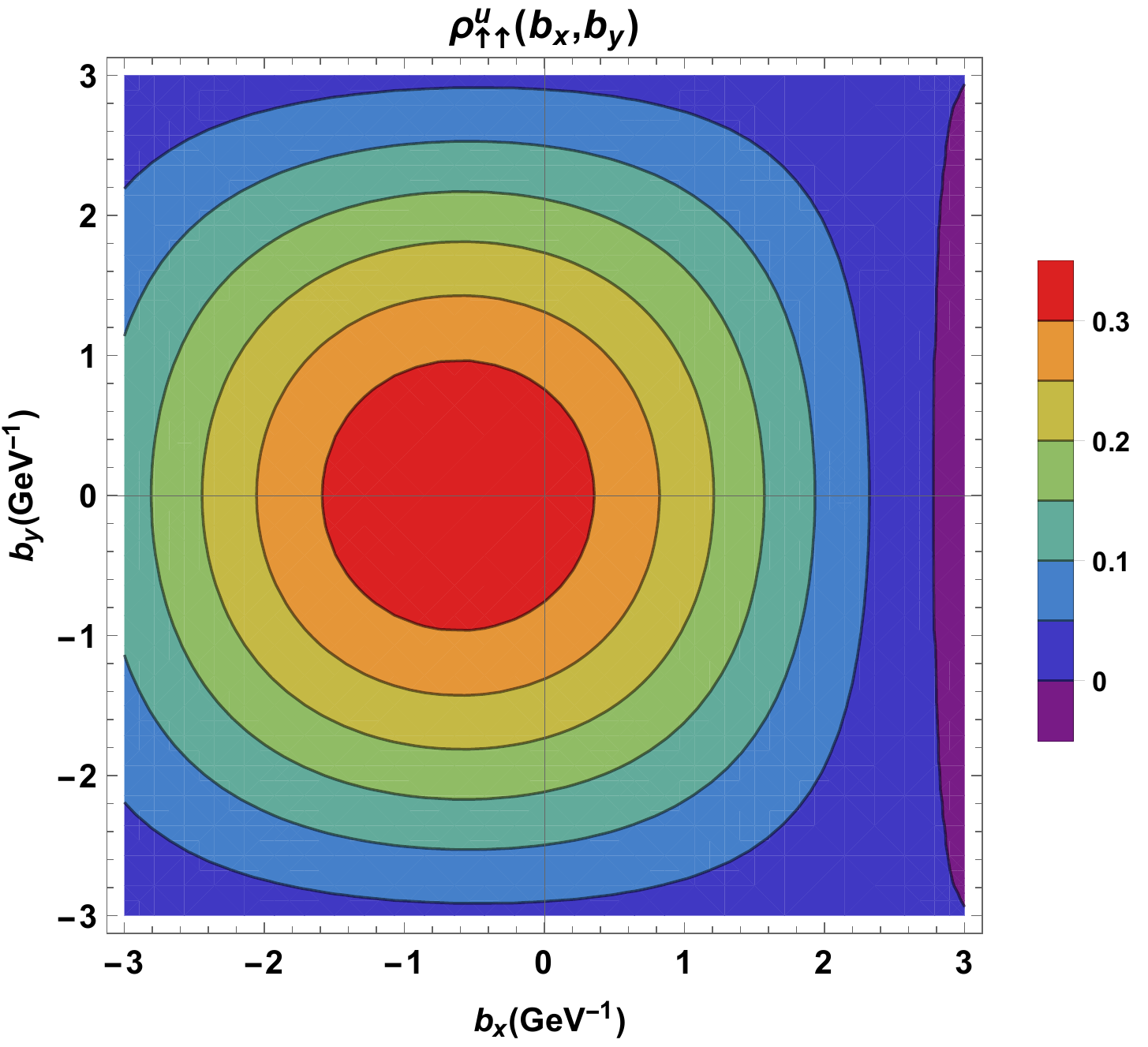}
(b)\includegraphics[width=.4\textwidth]{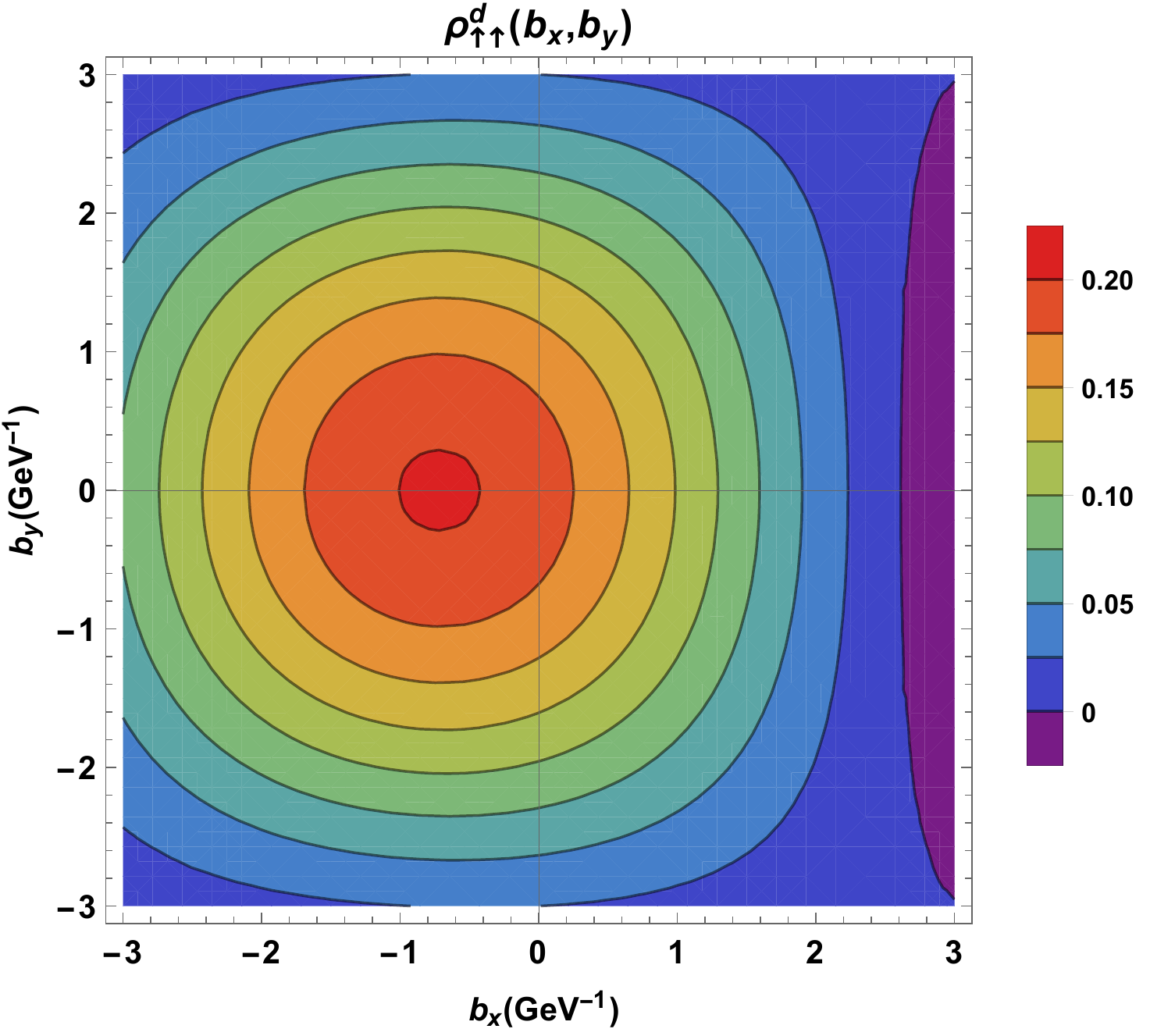}
\end{minipage}
\begin{minipage}[c]{1\textwidth}
(c)\includegraphics[width=.4\textwidth]{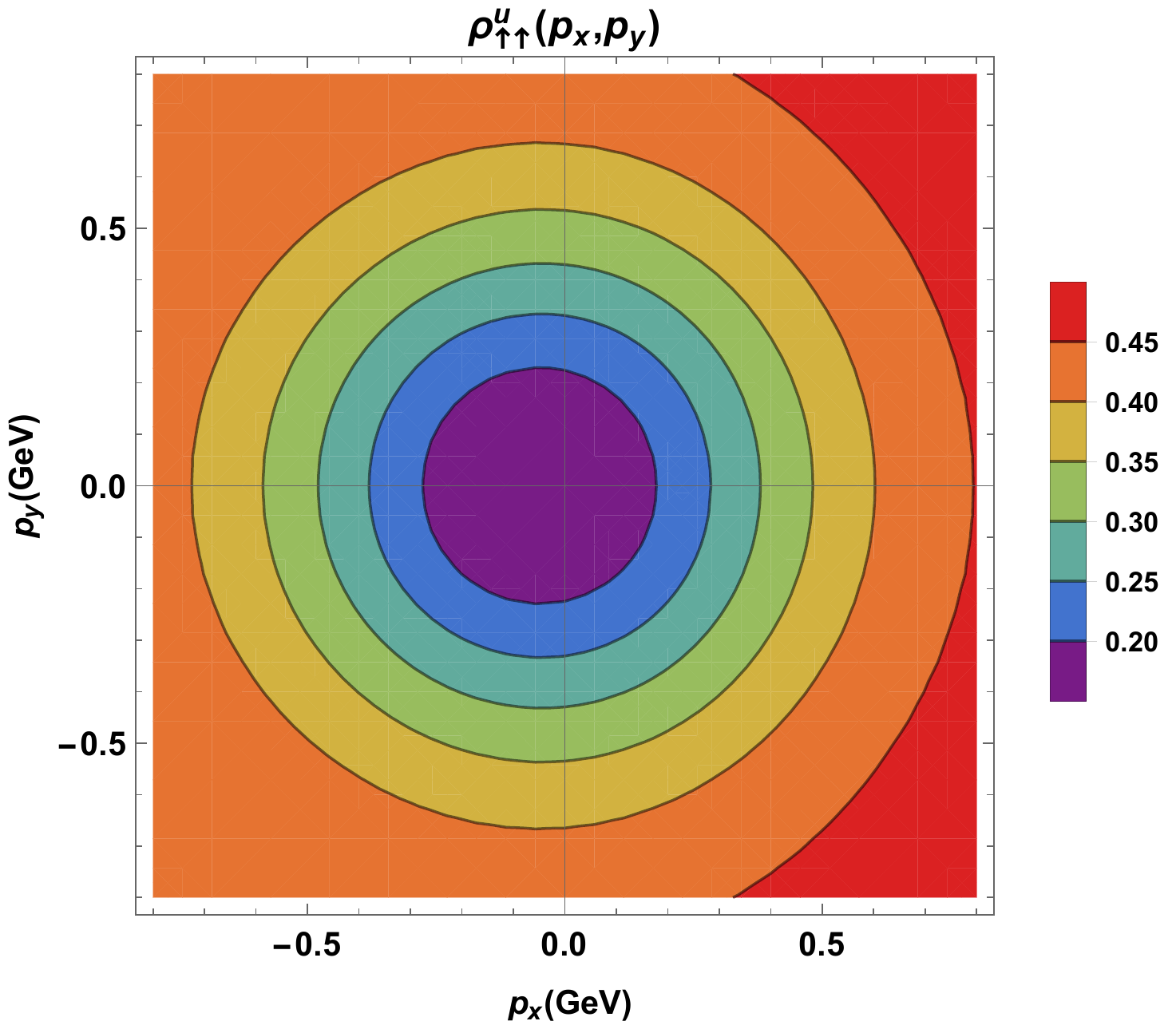}
(d)\includegraphics[width=.4\textwidth]{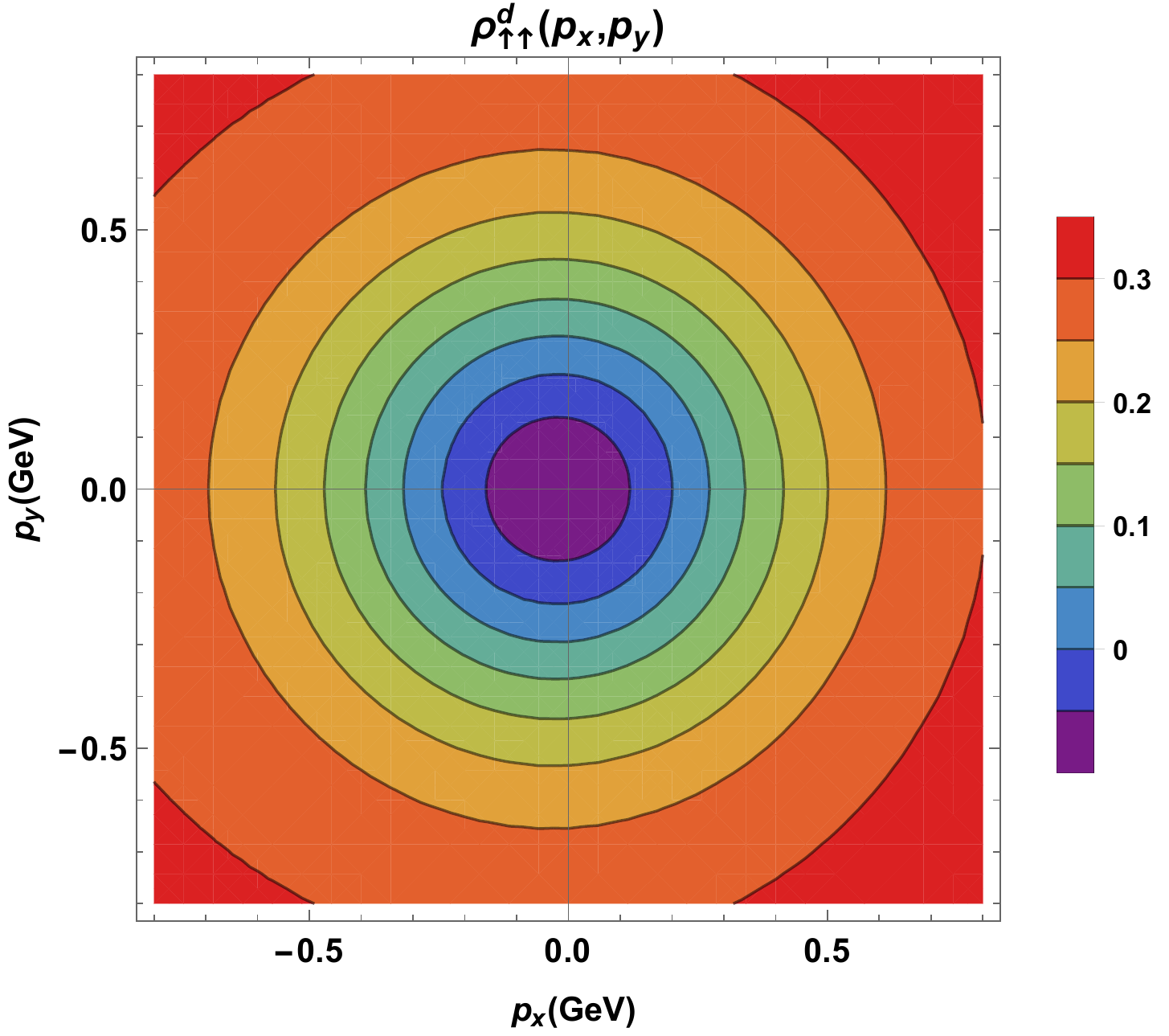}
\end{minipage}
\caption{The plot of Wigner distribution $\rho_{\uparrow \uparrow}({\bf b}_\perp,{\bf p}_\perp)$ in transverse impact-parameter plane and transverse momentum plane for $u$-quark (left panel) and $d$-quark (right panel).}
\label{longi-up-up}
\end{figure}
\begin{figure}
\centering
\begin{minipage}[c]{1\textwidth}
(a)\includegraphics[width=.4\textwidth]{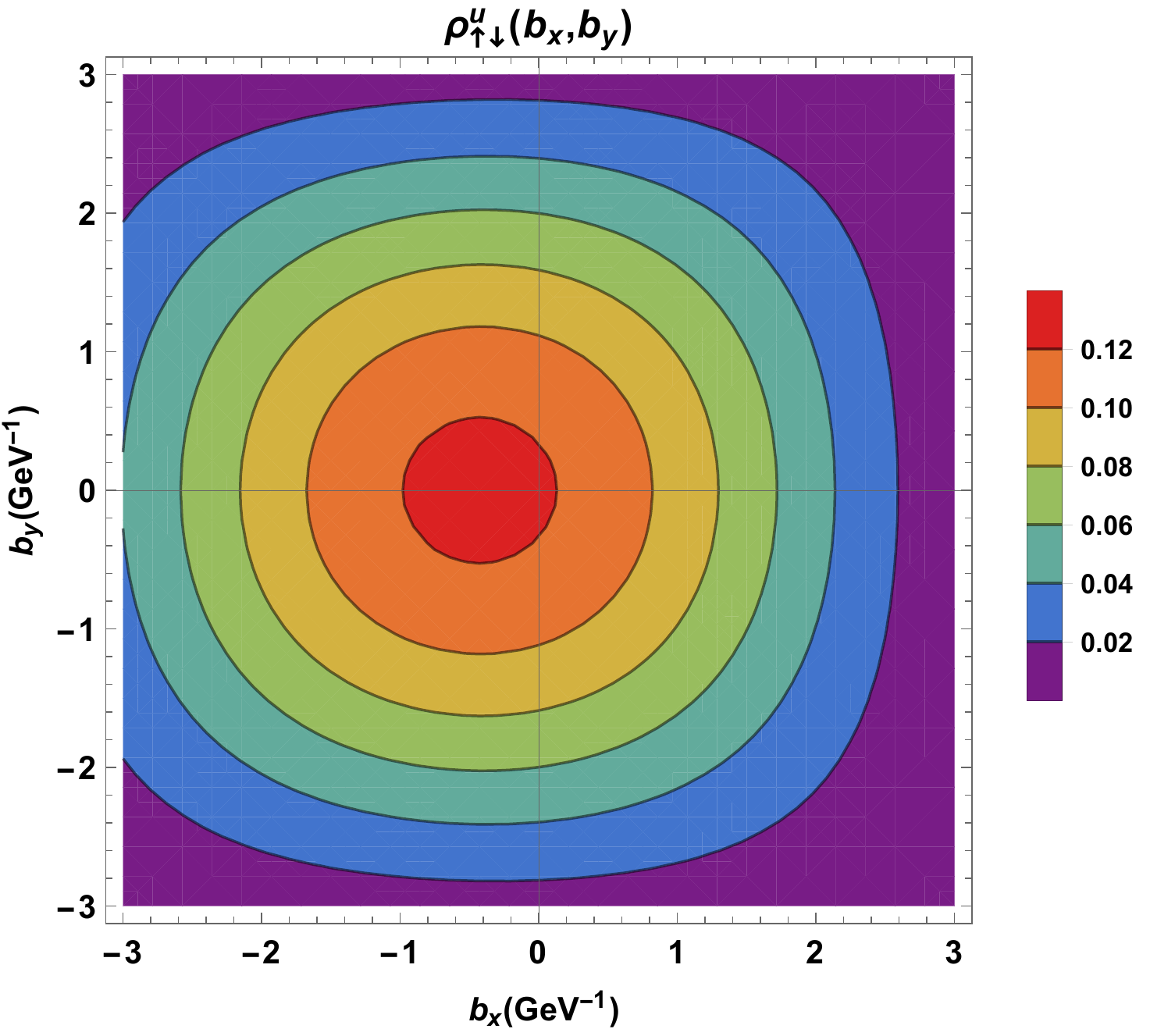}
(b)\includegraphics[width=.4\textwidth]{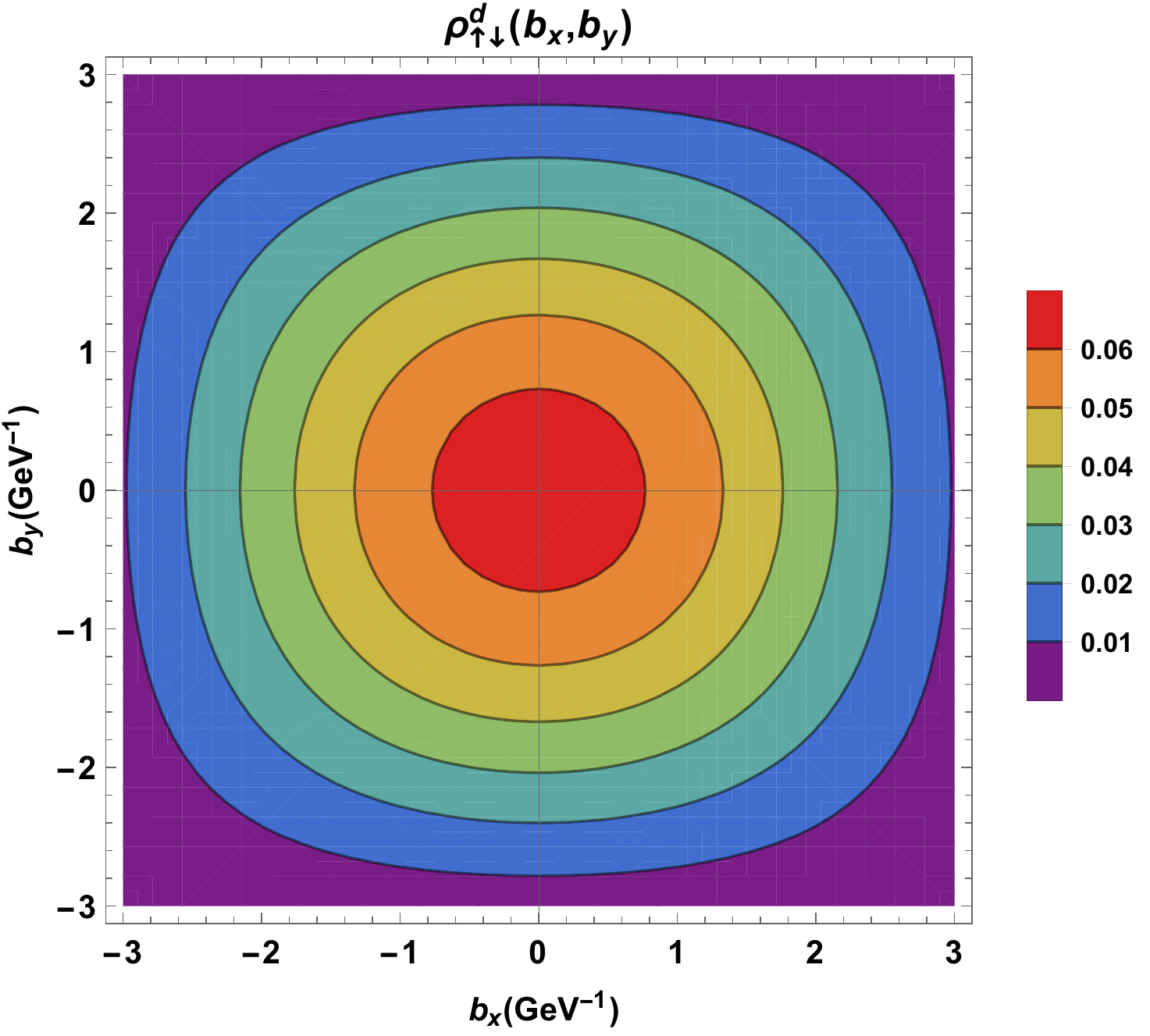}
\end{minipage}
\begin{minipage}[c]{1\textwidth}
(c)\includegraphics[width=.4\textwidth]{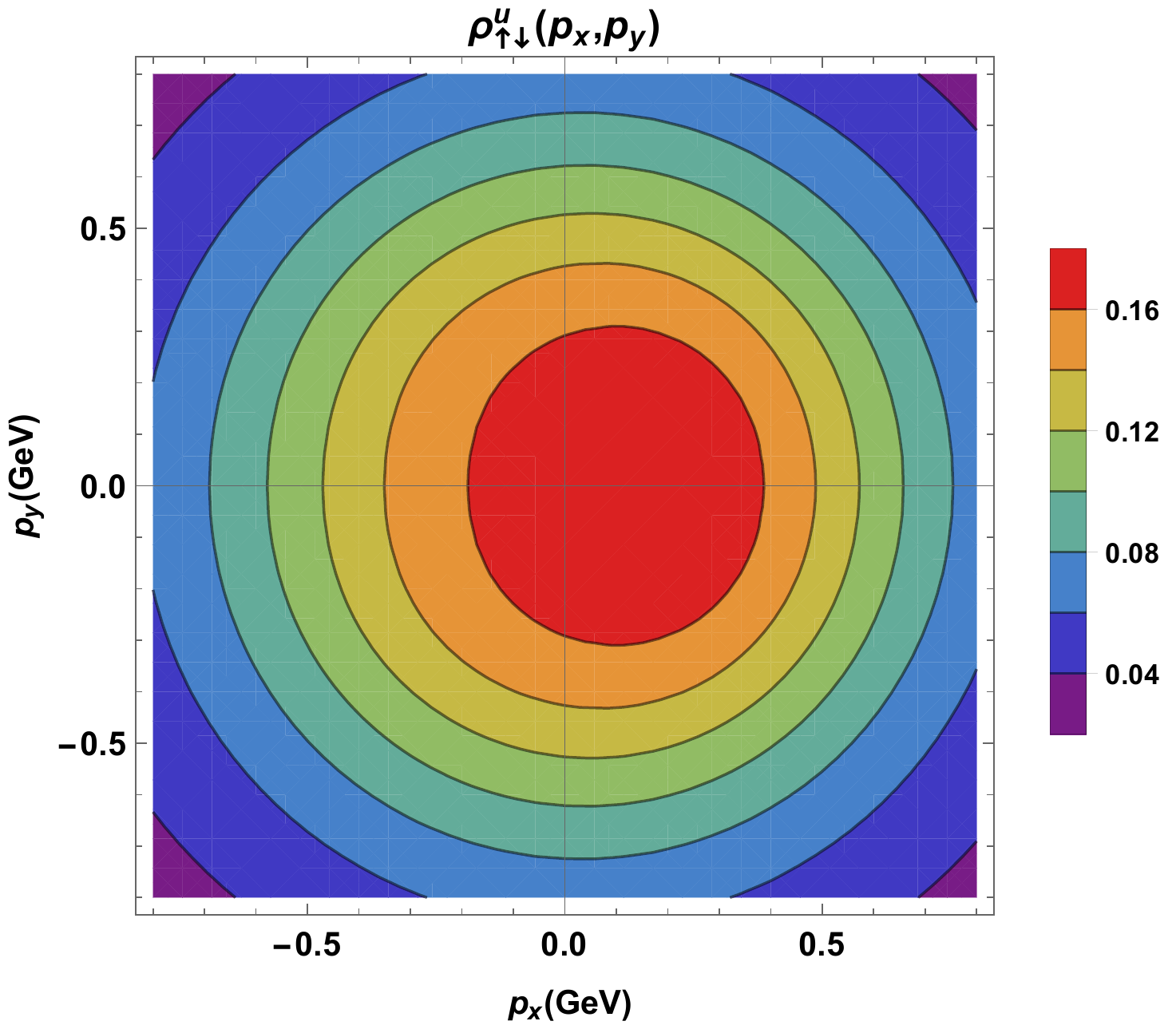}
(d)\includegraphics[width=.4\textwidth]{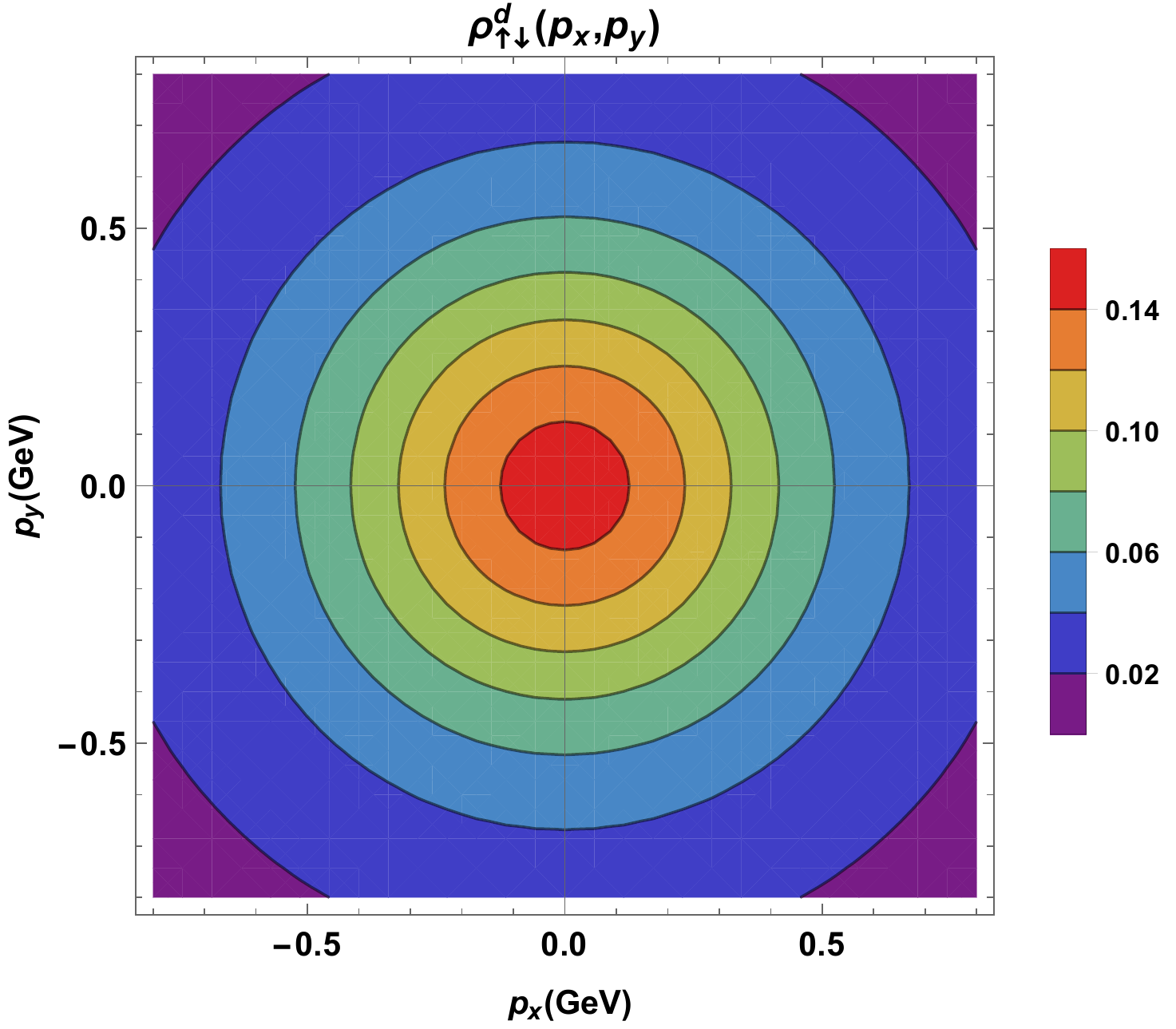}
\end{minipage}
\caption{The plot of Wigner distribution $\rho_{\uparrow \downarrow}({\bf b}_\perp,{\bf p}_\perp)$ in transverse impact-parameter plane and transverse momentum plane for $u$-quark (left panel) and $d$-quark (right panel).}
\label{longi-up-down}
\end{figure}
\begin{figure}
\centering
\begin{minipage}[c]{1\textwidth}
(a)\includegraphics[width=.4\textwidth]{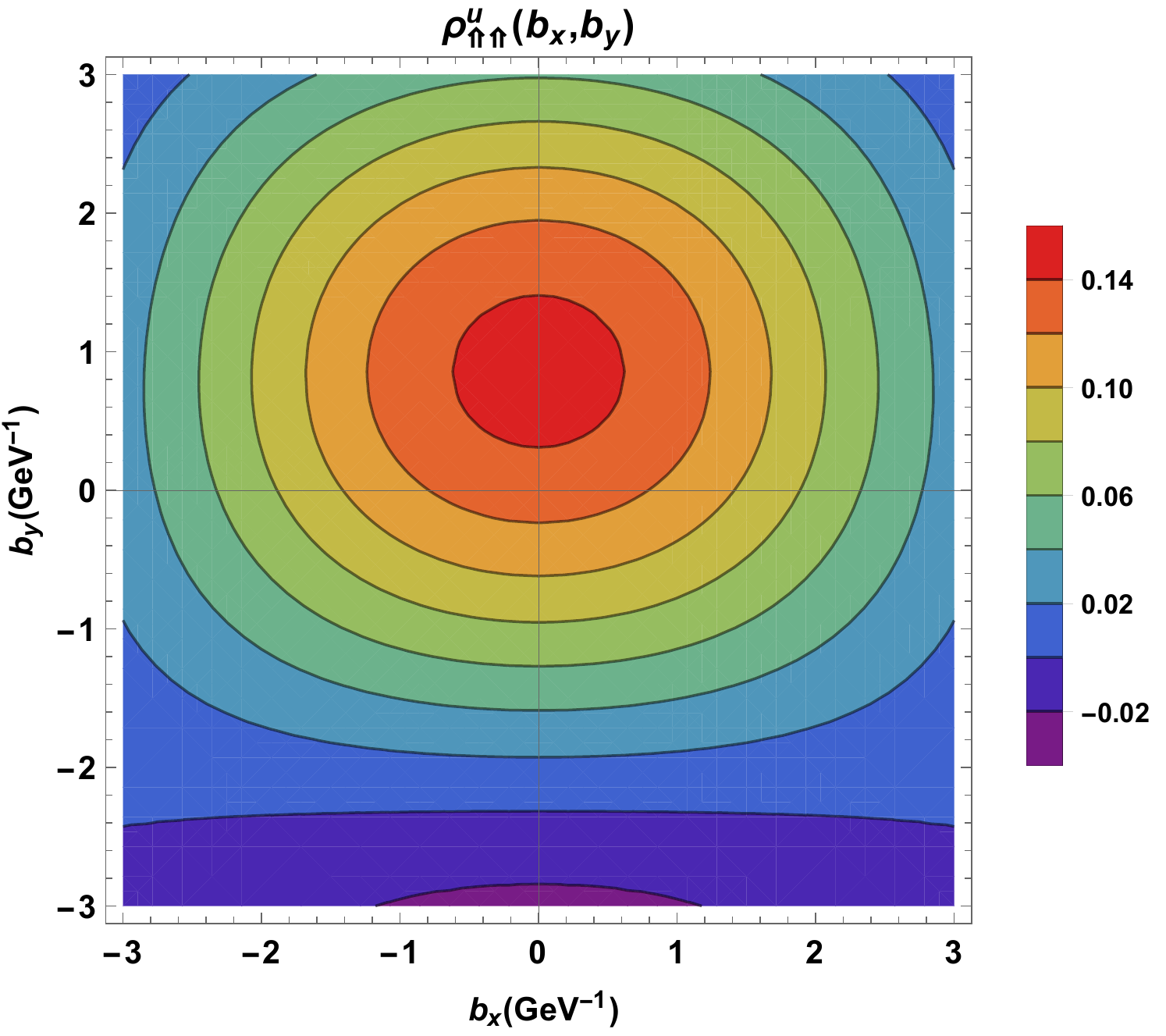}
(b)\includegraphics[width=.4\textwidth]{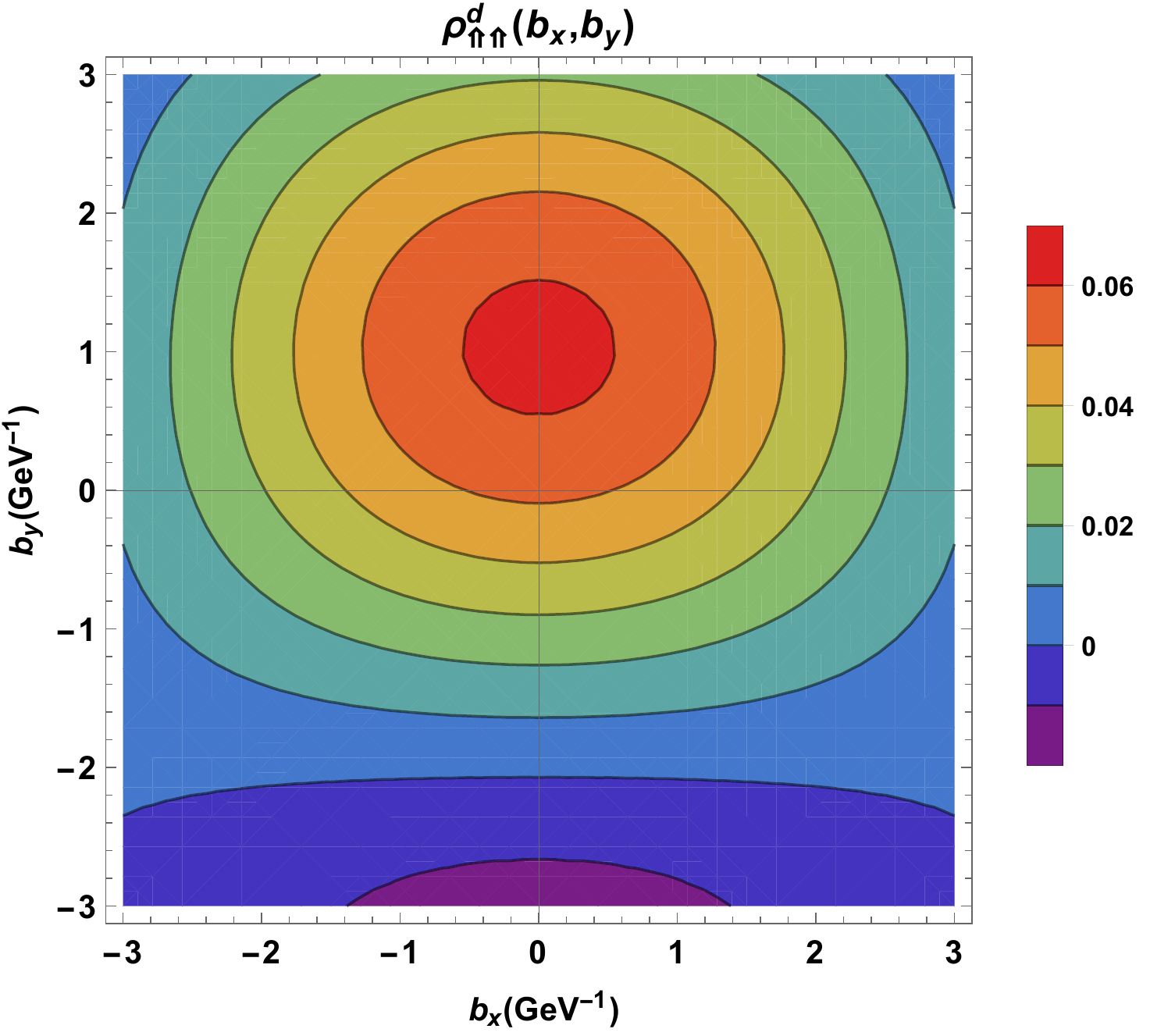}
\end{minipage}
\begin{minipage}[c]{1\textwidth}
(c)\includegraphics[width=.4\textwidth]{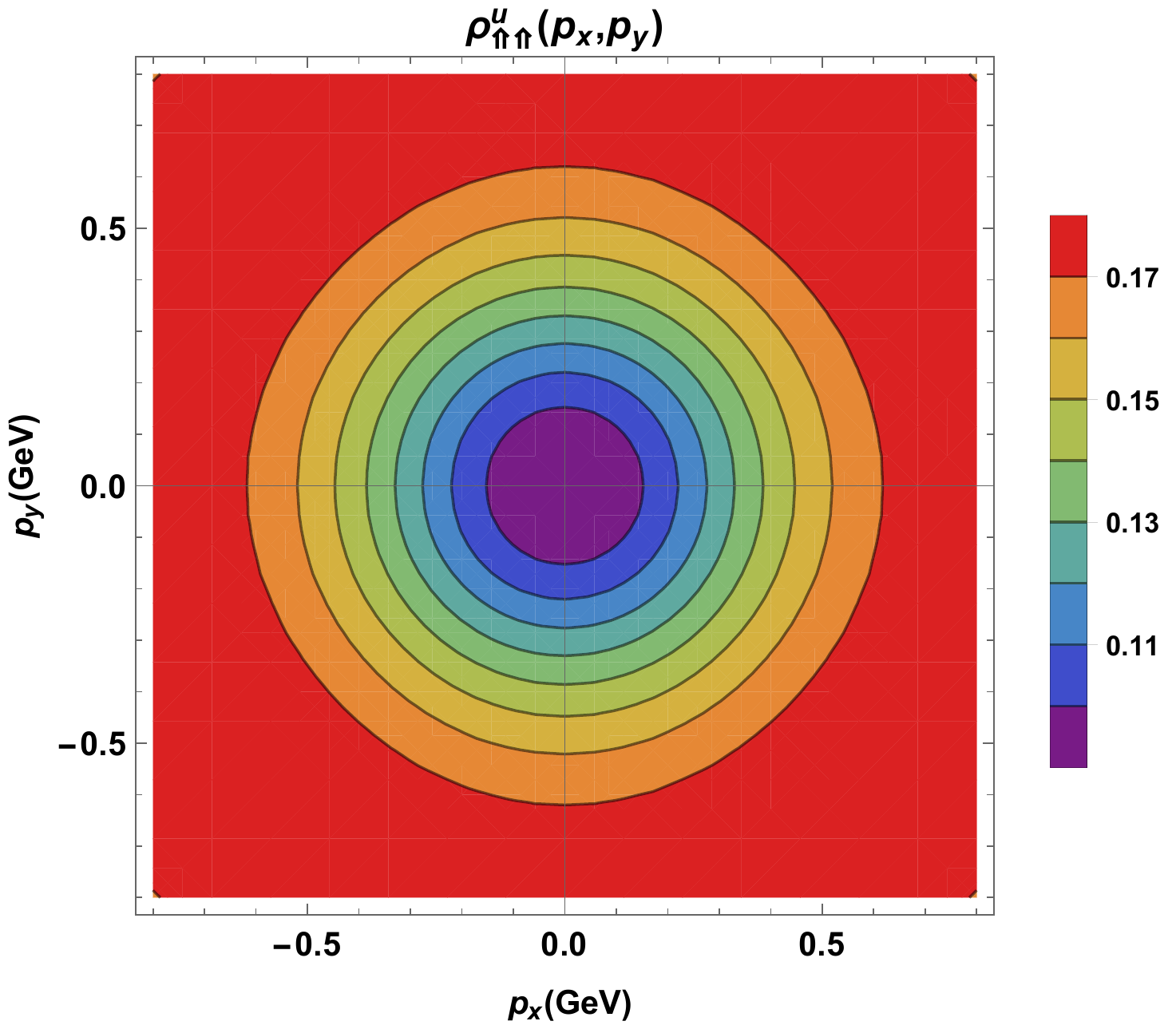}
(d)\includegraphics[width=.4\textwidth]{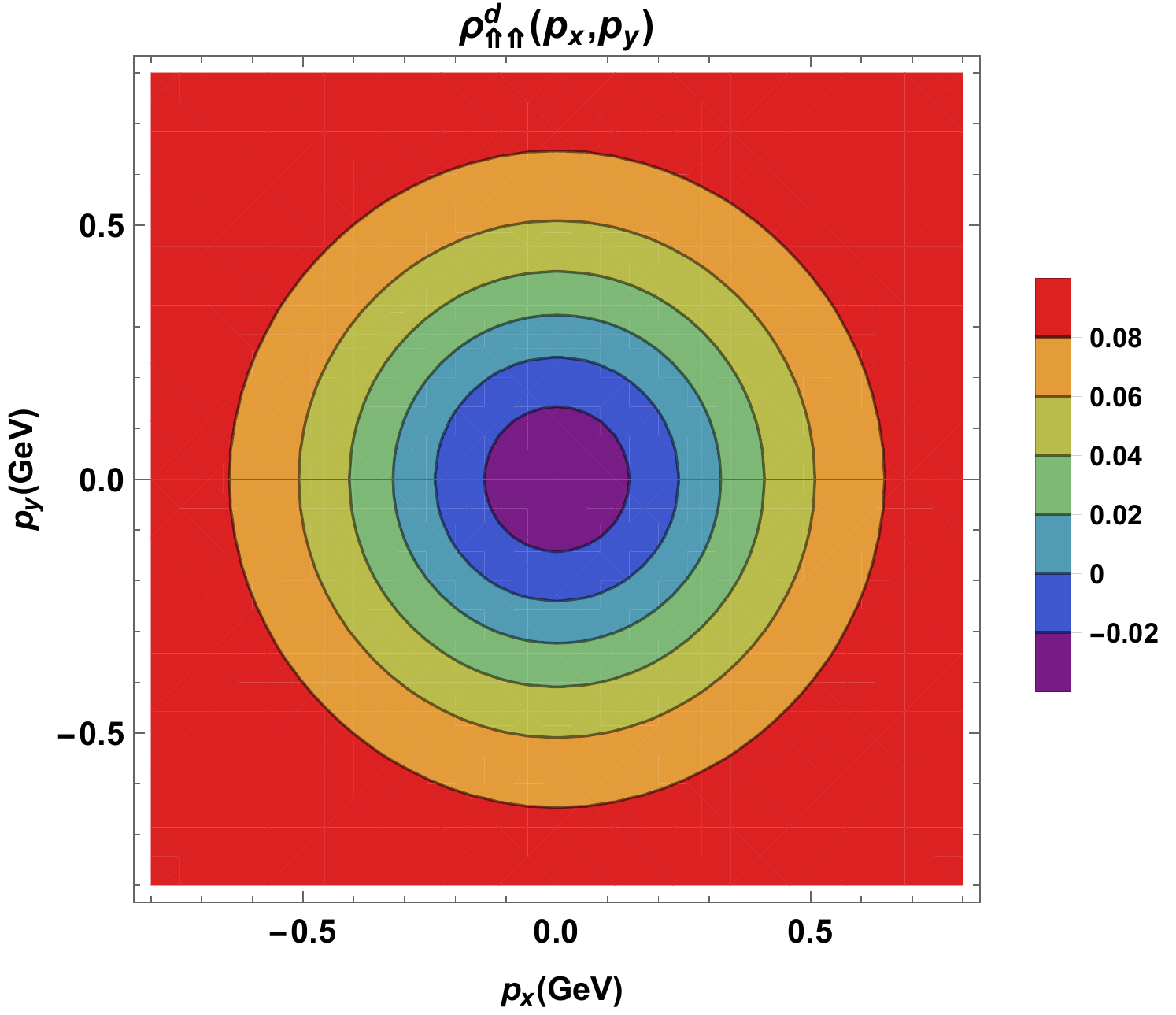}
\end{minipage}
\caption{The plot of Wigner distribution $\rho_{\Uparrow \Uparrow}({\bf b}_\perp,{\bf p}_\perp)$ in transverse impact-parameter plane and transverse momentum plane for $u$-quark (left panel) and $d$-quark (right panel).}
\label{trans-up-up}
\end{figure}
\begin{figure}
\centering
\begin{minipage}[c]{1\textwidth}
(a)\includegraphics[width=.4\textwidth]{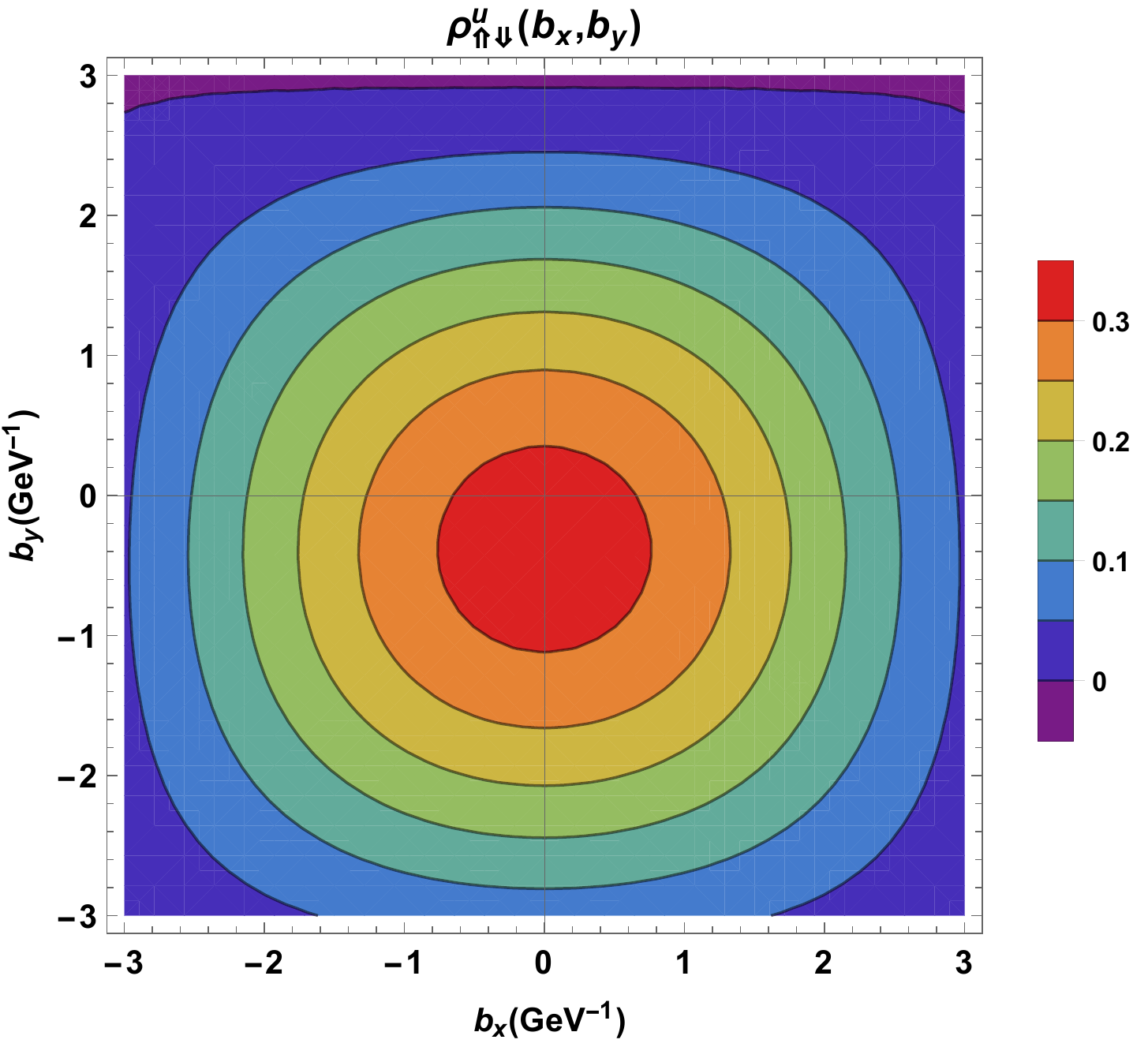}
(b)\includegraphics[width=.4\textwidth]{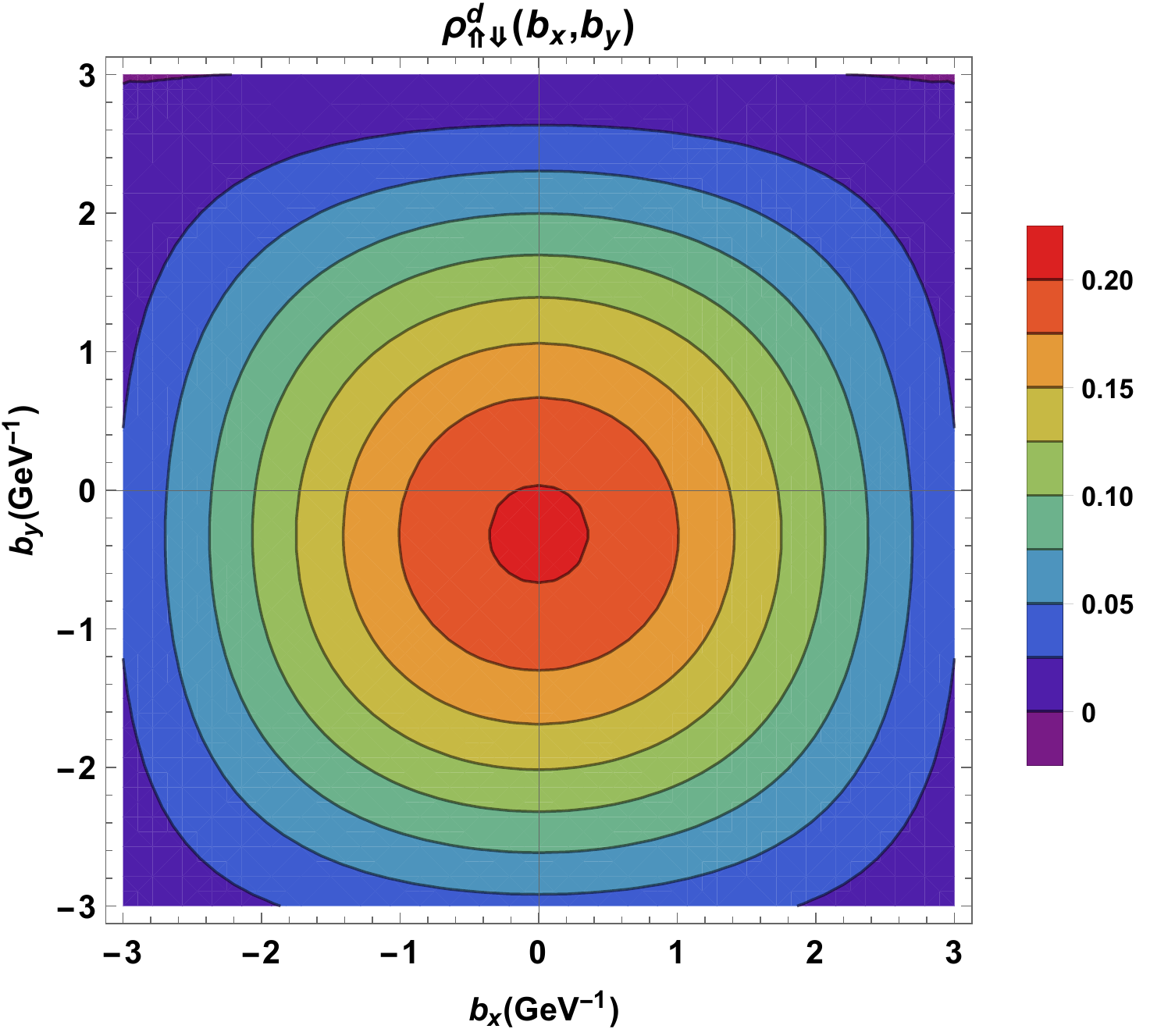}
\end{minipage}
\begin{minipage}[c]{1\textwidth}
(c)\includegraphics[width=.4\textwidth]{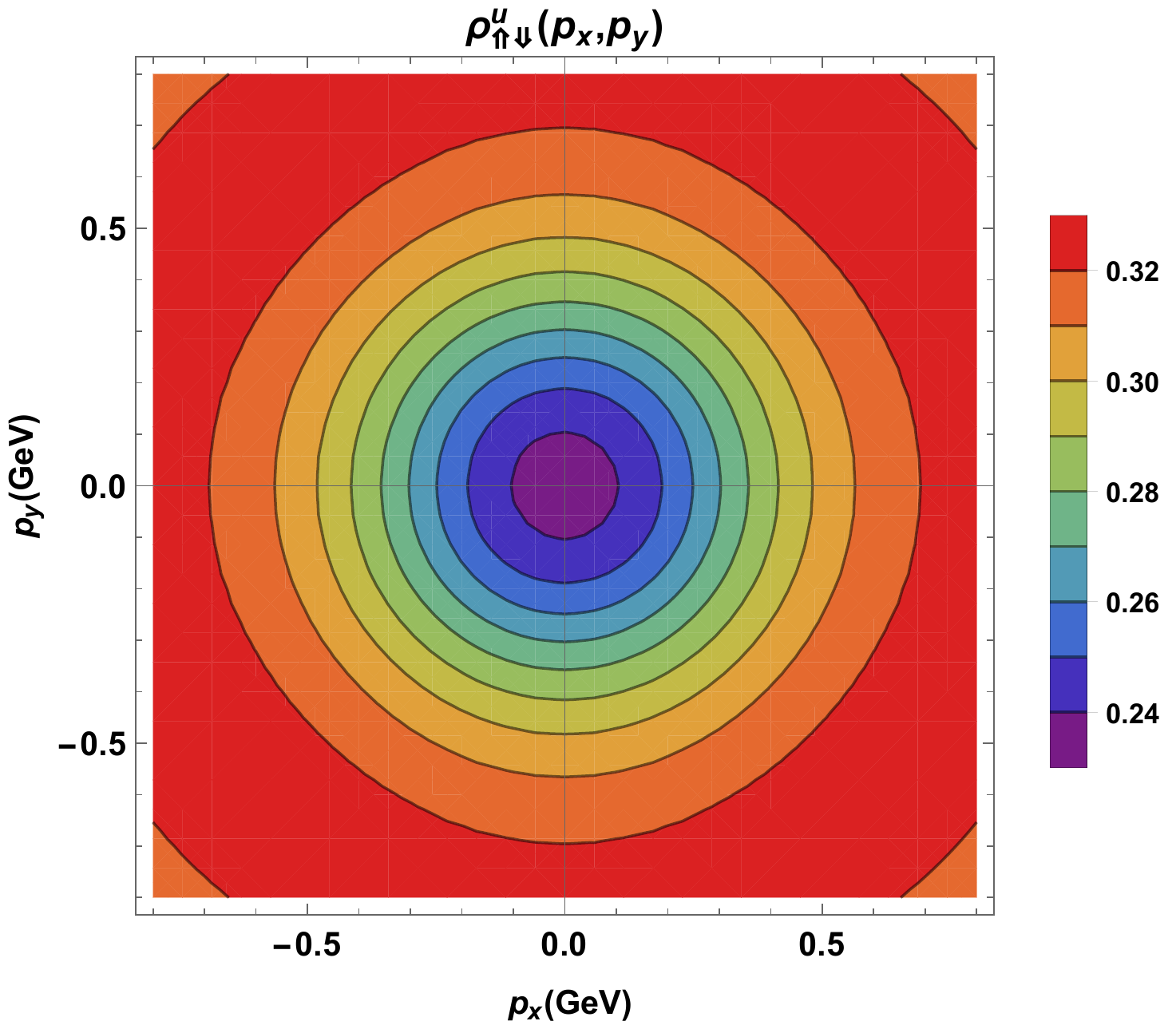}
(d)\includegraphics[width=.4\textwidth]{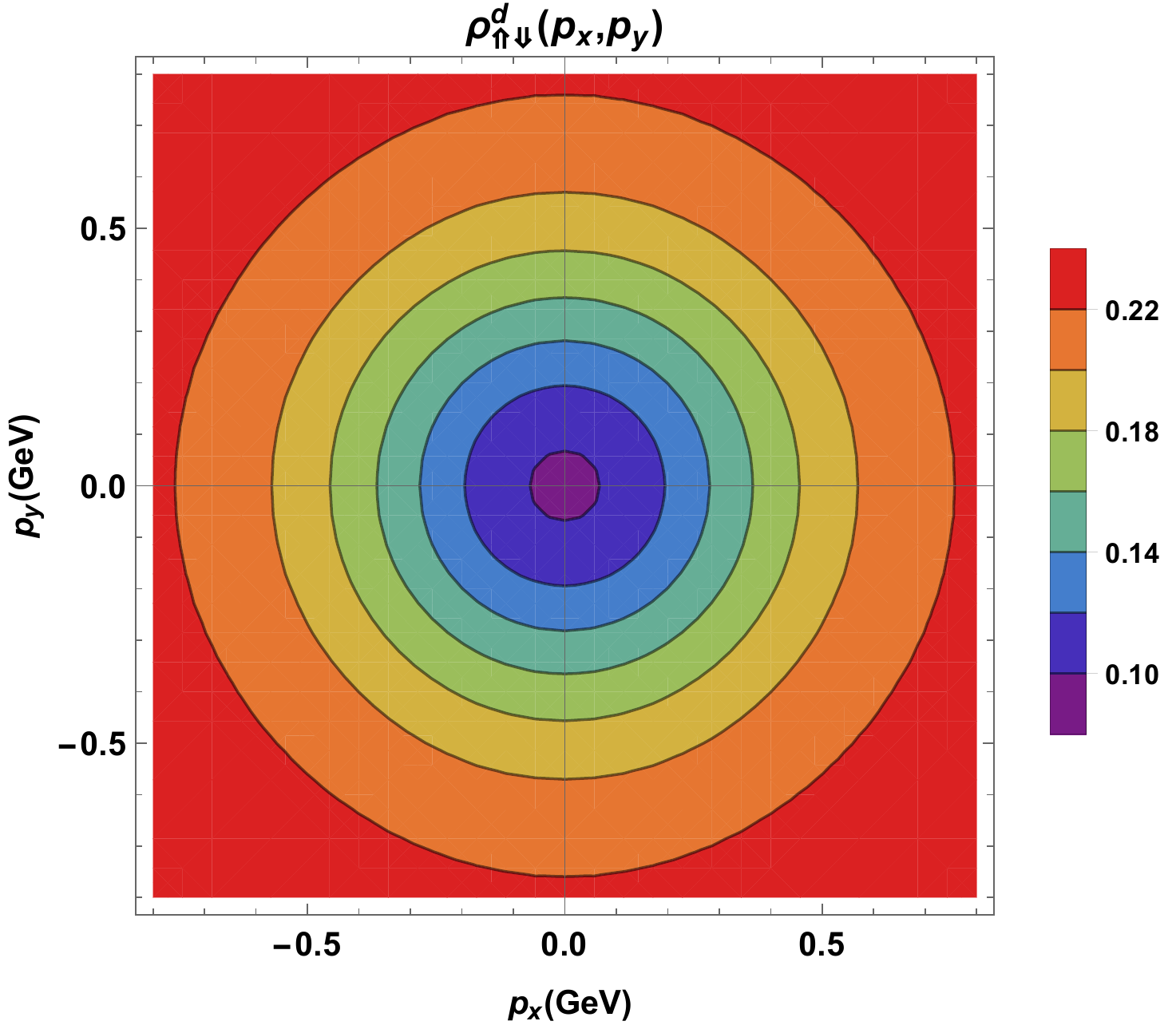}
\end{minipage}
\caption{The plot of Wigner distribution $\rho_{\Uparrow \Downarrow}({\bf b}_\perp,{\bf p}_\perp)$ in transverse impact-parameter plane and transverse momentum plane for $u$-quark (left panel) and $d$-quark (right panel).}
\label{trans-up-down}
\end{figure}
\begin{figure}
\centering
\begin{minipage}[c]{1\textwidth}
(a)\includegraphics[width=.4\textwidth]{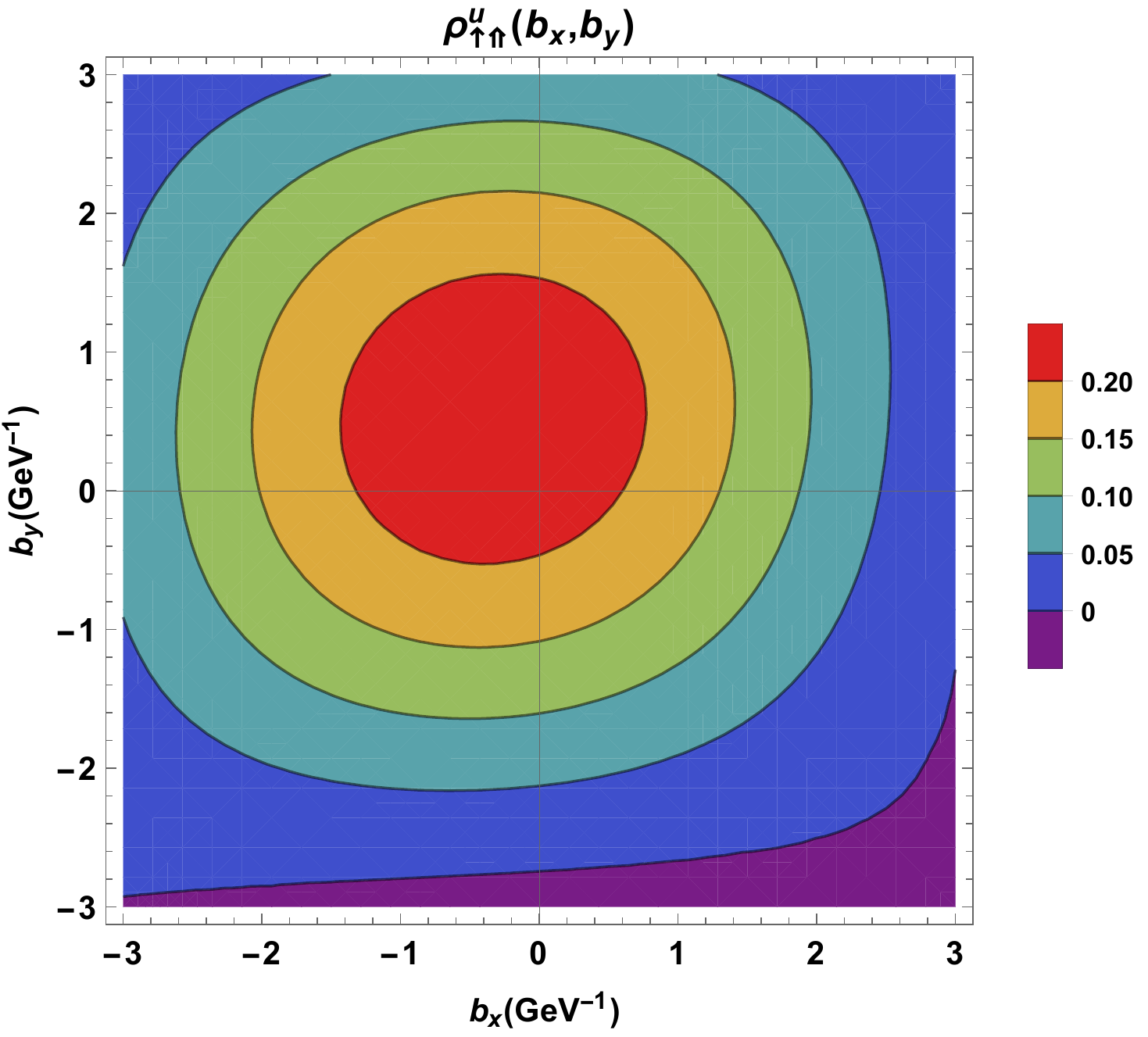}
(b)\includegraphics[width=.4\textwidth]{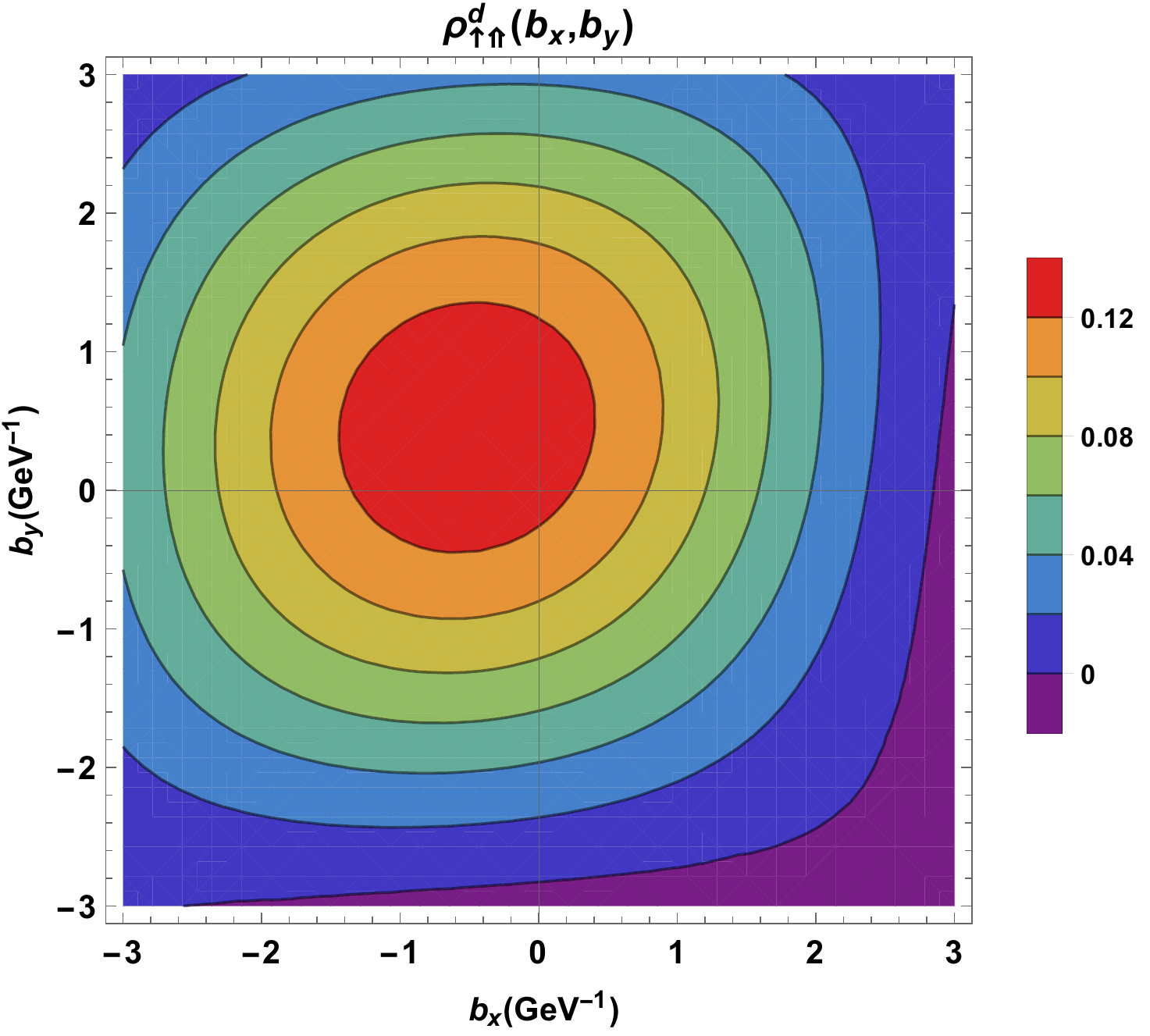}
\end{minipage}
\begin{minipage}[c]{1\textwidth}
(c)\includegraphics[width=.4\textwidth]{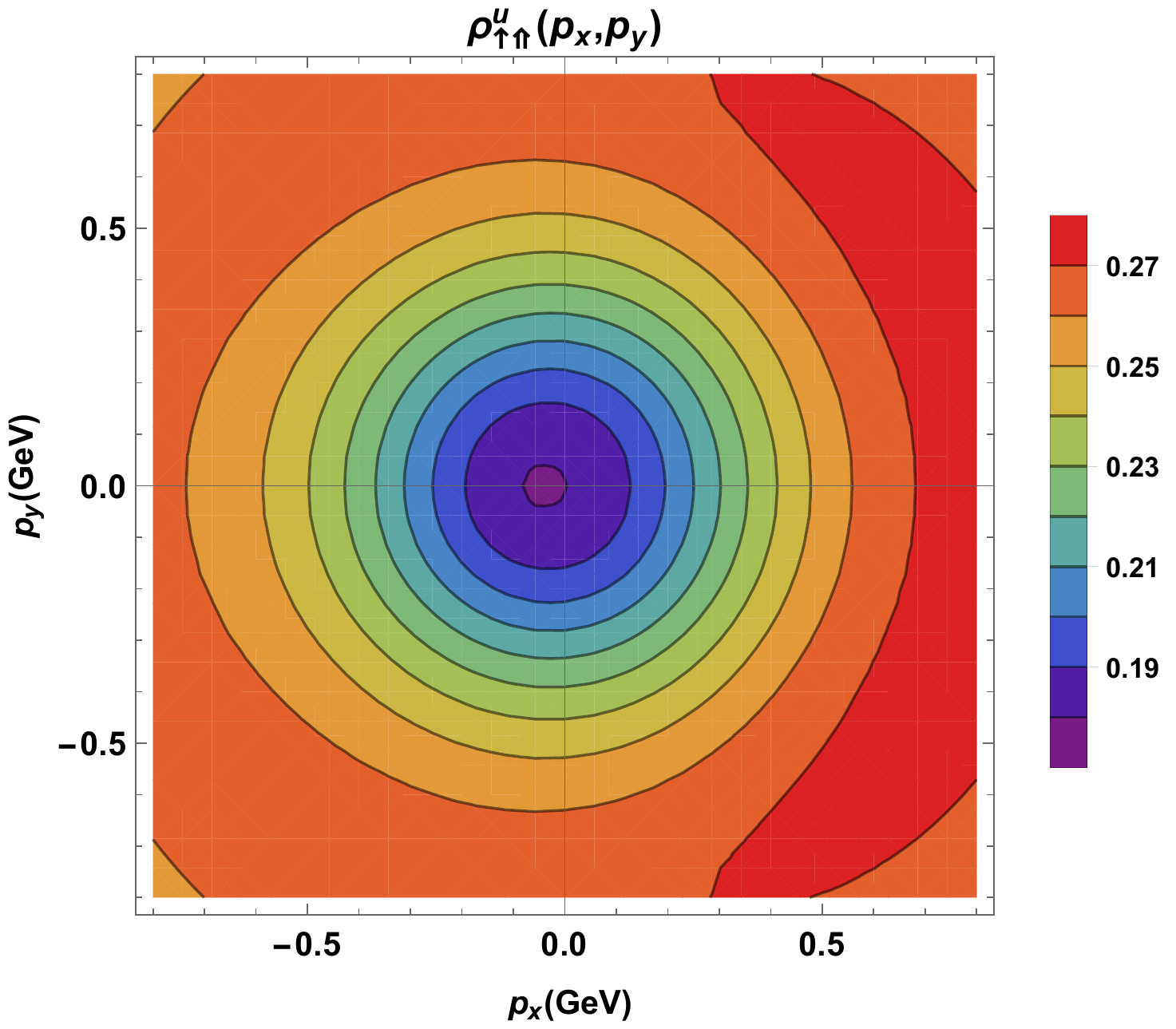}
(d)\includegraphics[width=.4\textwidth]{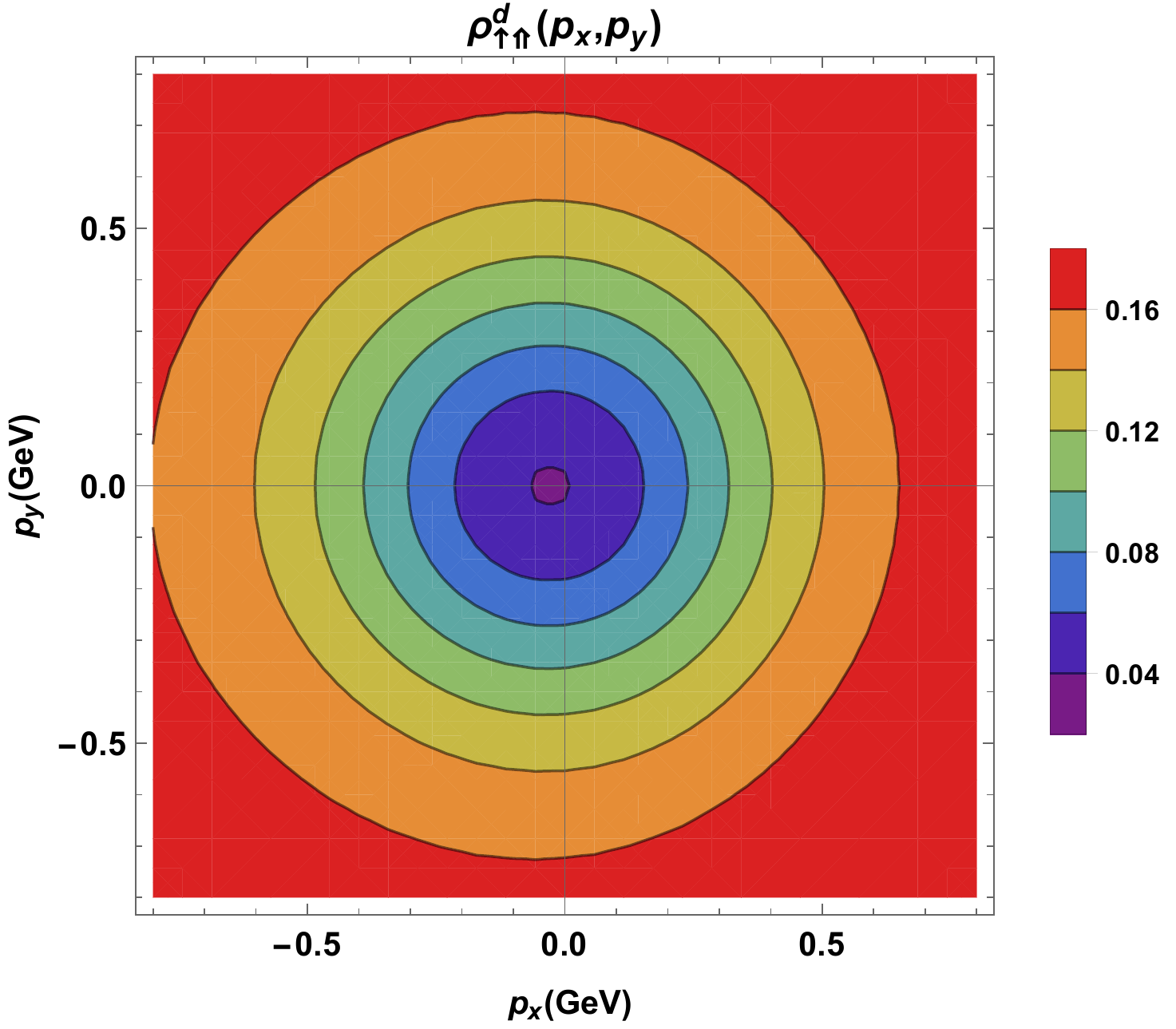}
\end{minipage}
\caption{The plot of Wigner distribution $\rho_{\uparrow \Uparrow}({\bf b}_\perp,{\bf p}_\perp)$ in transverse impact-parameter plane and transverse momentum plane for $u$-quark (left panel) and $d$-quark (right panel).}
\label{longi-up-trans-up}
\end{figure}
Further, the correlator $W_{XY}$ is related to the Wigner distribution as 
\begin{eqnarray}
\rho_{XY}({\bf b}_\perp, {\bf p}_\perp, x,S)=\int \frac{d^2 {\bf \Delta}_\perp}{(2\pi)^2}e^{-i{\bf \Delta}_\perp . {\bf b}_\perp } W_{XY}({\bf \Delta}_\perp, {\bf p}_\perp, x,S),
\end{eqnarray} 
where $X$ and $Y$ being the composite particle and quark polarizations.

In this work, the DGLAP region for quarks is used to evaluate the Wigner distributions i.e. $0<x<1$. The respective momenta of initial and final state of struck quark in symmetric frame are defined as
\begin{eqnarray}
{\bf p}'_\perp&=&{\bf p}_\perp-(1-x)\frac{{\bf \Delta}_\perp}{2},\nonumber\\
{\bf p}''_\perp&=&{\bf p}_\perp+(1-x)\frac{{\bf \Delta}_\perp}{2}.
\end{eqnarray}

The superposition of scalar and axial-vector diquark results into the quark flavors as \cite{tmd-bachhetta}
\begin{eqnarray}
\rho^u&=&c_s^2 \rho^{u(s)}+c_a^2 \rho^{u(a)},\nonumber\\
\rho^d&=&c_a'^2 \rho^{d(a')},
\end{eqnarray}
where the superscripts $(s)$, $(a)$ and $(a')$ denote the scalar isoscalar,  vector-isoscalar and vector isovector diquarks respectively. The mass values and couplings for diquarks  have been summarized in Table \ref{table1}.

\begin{table}[ht]
\begin{center}
\begin{tabular}{||@{\hspace{10pt}} c @{\hspace{10pt}} ||
@{\hspace{10pt}} c @{\hspace{10pt}} | @{\hspace{10pt}} c
@{\hspace{10pt}}| @{\hspace{10pt}} c @{\hspace{10pt}}||}
%
%\hline\hline
%$Diquark$ & {$M_X$ in $GeV$ {\hspace{9pt}} } & {$c_X$ {\hspace{9pt}}} \\
%$ud$ (Scalar $s$) & 0.822 $\pm$ 0.053 & 0.847 $\pm$ 0.111 \\
%$ud$ (Axial-vector $a$) & 1.492 $\pm$ 0.173 & 1.061 $\pm$ 0.085 \\
%$uu$ (Axial-vector $a'$) & 0.890 $\pm$ 0.008 & 0.880 $\pm$ 0.008 \\
\hline
 $Diquark$ & $ud$ (Scalar $s$) & $ud$ (Axial-vector $a$) & $uu$ (Axial-vector $a'$) \\
 \hline
{$M_X$ in $GeV$ {\hspace{9pt}} } & 0.822 $\pm$ 0.053 & 1.492 $\pm$ 0.173 &  0.890 $\pm$ 0.008 \\
\hline
 {$c_X$ {\hspace{9pt}}} & 0.847 $\pm$ 0.111 & 1.061 $\pm$ 0.085 & 0.880 $\pm$ 0.008 \\
\hline 
\end{tabular}
\caption{The diquark masses $M_X$ and couplings $c_X$ for the scalar isoscalar, vector  isoscalar diquark and vector isovector diquark.}
\label{table1}
\end{center}
\end{table}

We plot the Wigner distributions of the quark in the fermion system having spins in longitudinal direction, i.e. $\rho_{\Lambda \lambda}$. Here, we take two cases for the discussion on the longitudinal Wigner distributions: (i) spin direction of composite system and quark to be $\Lambda=\uparrow$ and $\lambda=\uparrow$ i.e. $\rho_{\uparrow \uparrow}$, (ii) proton polarization $\Lambda=\uparrow$ and quark polarization $\lambda=\downarrow$ i.e. $\rho_{\uparrow \downarrow}$.  In Fig. \ref{longi-up-up} (a) and (b), we plot the longitudinal distribution $\rho_{\uparrow \uparrow}$ for $u$-quark and $d$-quark respectively. We see the distribution effects in transverse impact-parameter plane and transverse momentum plane. The distribution shows circular behaviour with the peaks shifting towards $b_x<0$ and $p_x<0$ in impact-parameter plane and momentum plane respectively. In momentum plane, as shown in Fig. \ref{longi-up-up} (c) and (d), we observe the distortion along $b_x$ at the higher values of impact-parameter co-ordinate. In this model, the distributions $\rho_{UL}$ and $\rho_{LU}$ are same for axial-vector diquark. Based on the different combinations of helicities, we get the different cases from Eq. (\ref{wigner_correlator}) as follows, \\
for scalar diquark 
\begin{eqnarray}
\rho_{\uparrow \uparrow}&=&\frac{1}{2}[\rho_{UU}+\rho_{LL}],\label{scalar-up-up}\nonumber\\
\rho_{\uparrow \downarrow}&=&\frac{1}{2}[\rho_{UU}-2\rho_{UL}-\rho_{LL}],
\label{scalar-up-down}
\end{eqnarray}
for axial-vector diquark
\begin{eqnarray}
\rho_{\uparrow \uparrow}&=&\frac{1}{2}[\rho_{UU}+2 \rho_{UL} +\rho_{LL}], \label{vector-up-up}\nonumber\\
\rho_{\uparrow \downarrow}&=&\frac{1}{2}[\rho_{UU}-\rho_{LL}].\label{vector-up-down} 
\end{eqnarray}
We plot the quark Wigner distribution having respective longitudinal polarization of quark $\lambda=\downarrow$ and fermion system $\Lambda=\uparrow$ in Fig. \ref{longi-up-down}. The distortion is observed in impact-parameter plane which gets more noticeable at the increasing values of ${\bf b}_\perp$ for $u$-quark and $d$-quark. The effect of distortion is more in case of $u$-quark as compared to $d$-quark. In momentum plane, the distortion is seen at the center of the $u$-quark distribution. The distribution plots look nearly similar for $\Lambda=\lambda$ and $\Lambda \neq \lambda$, as  Eqs. (\ref{scalar-up-up}) and (\ref{vector-up-down}) contribute the same terms. The polarities are opposite for distribution of $\Lambda=\lambda$ in ${\bf p}_\perp$-plane and ${\bf b}_\perp$-plane. Since the distribution contributions from $\rho_{UU}$ and $\rho_{LL}$ are circularly symmetric (shown in Ref. \cite{wd8}), the distortion appears in the plots of $\rho_{\uparrow \uparrow}$ and $\rho_{\uparrow \downarrow}$ due to the addition of terms $\rho_{UL}$ and $\rho_{LU}$. In other words, the contribution from $\rho_{\uparrow \uparrow}$ is cirularly symmteric, because the interference of $\rho_{UL}$ and $\rho_{LU}$ is destructive, but when we add the axial vector part along with the scalar part to get the distribution of $u$-quark and $d$-quark in proton, the distortion takes place. Similar is the case of $\rho_{\uparrow \downarrow}$, however here the unpolarized-longitudinal Wigner distribution and longitudinal-unplarized Wigner distribution interfere destructively in axial-vector diquark case instead in scalar-diquark case, constructive interference is there. These interferences when added up accordingly, as Eqs. (\ref{scalar-up-down}) and (\ref{vector-up-down}), cause the sideway shifts of distributions as shown in Figs. \ref{longi-up-up} and \ref{longi-up-down}. 
\begin{figure}
\centering
\begin{minipage}[c]{1\textwidth}
(a)\includegraphics[width=.4\textwidth]{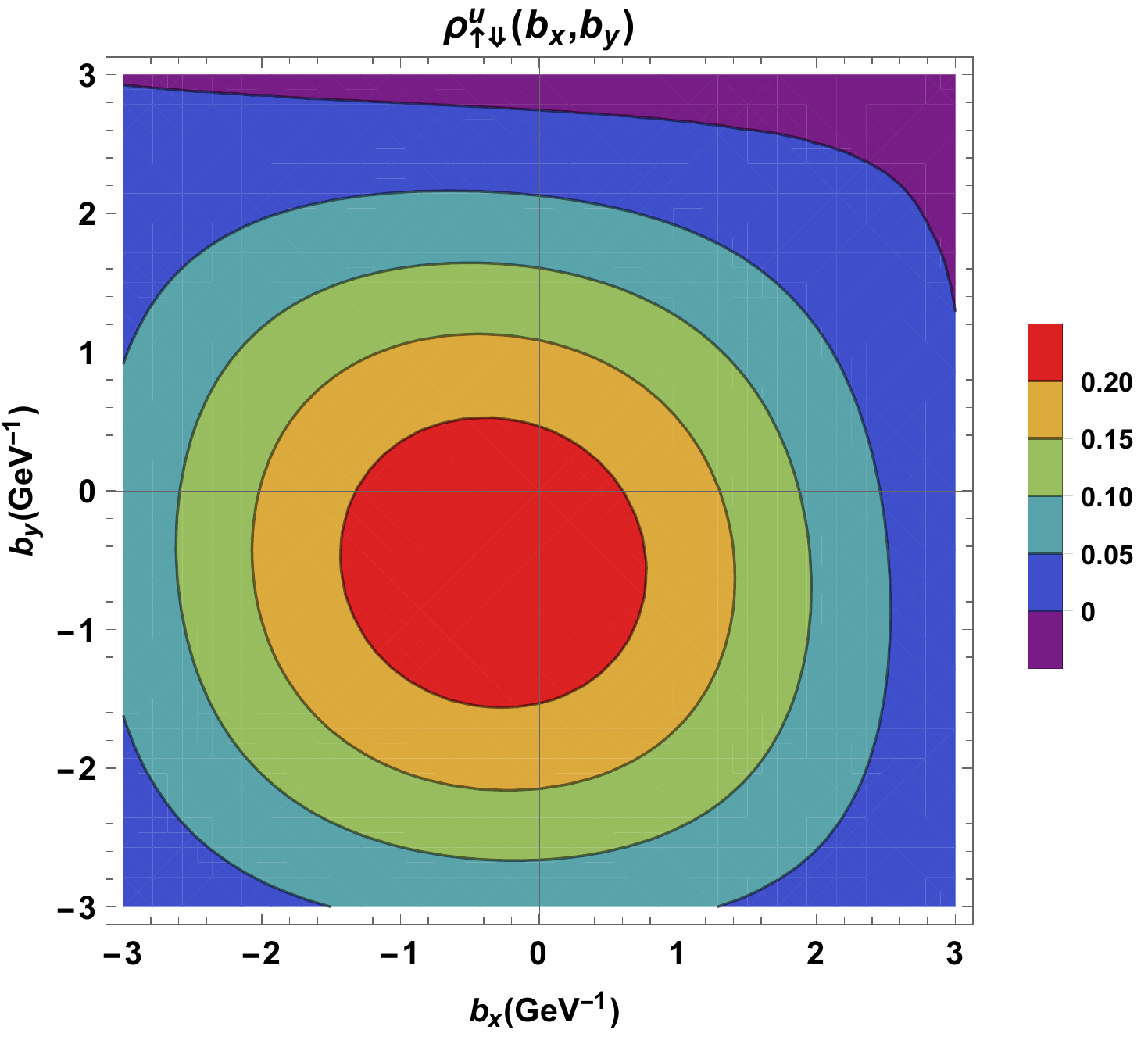}
(b)\includegraphics[width=.4\textwidth]{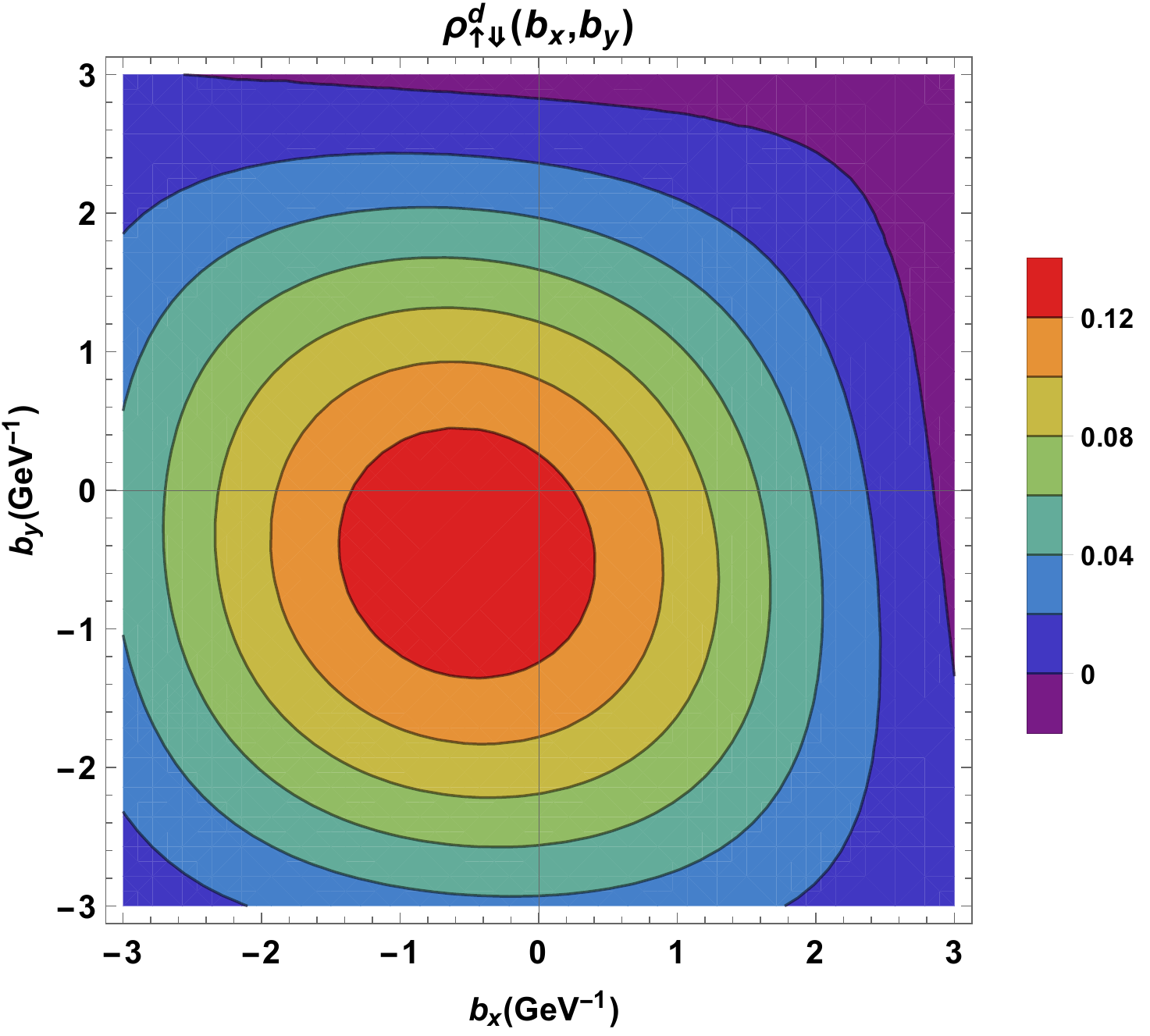}
\end{minipage}
\begin{minipage}[c]{1\textwidth}
(c)\includegraphics[width=.4\textwidth]{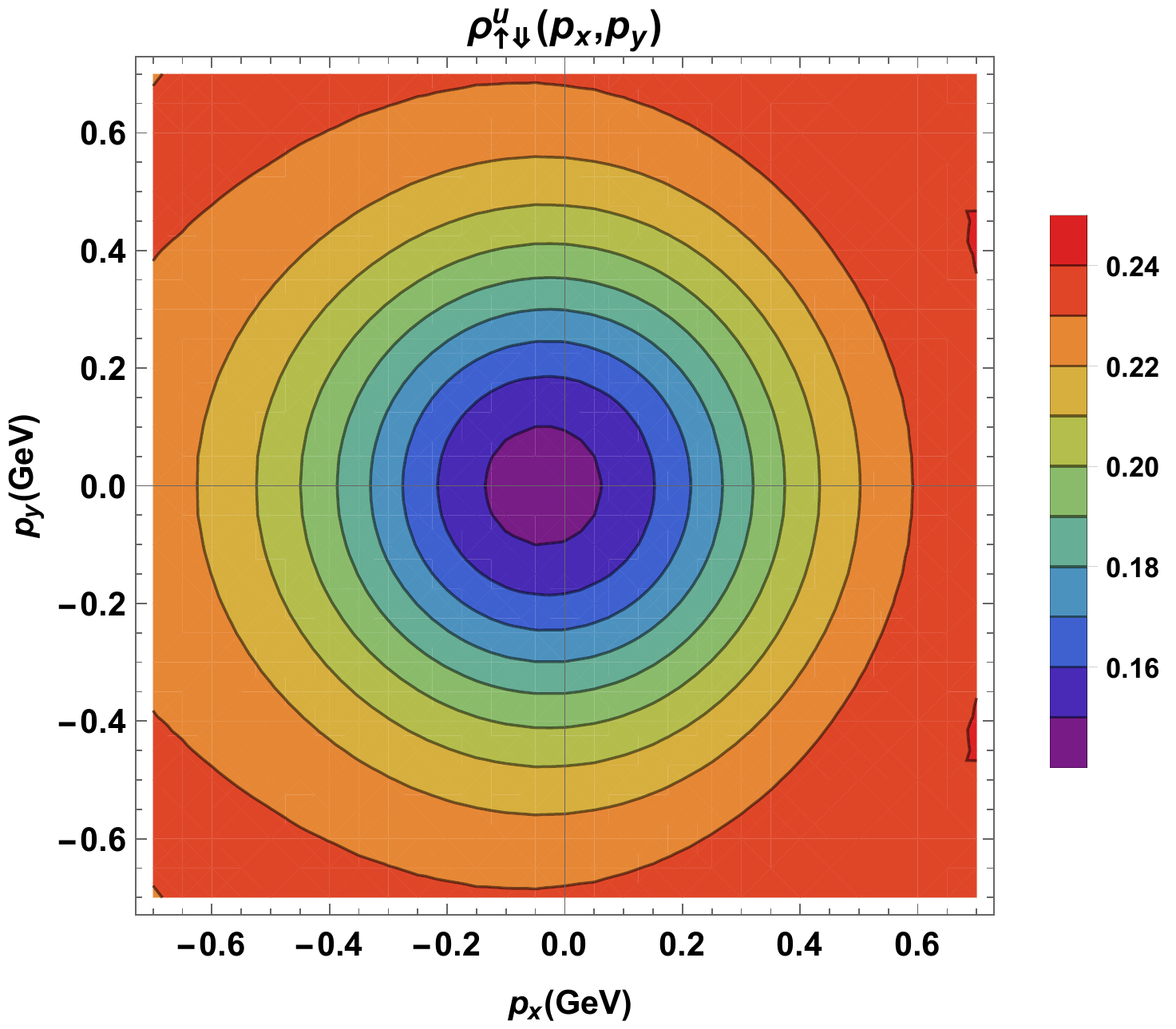}
(d)\includegraphics[width=.4\textwidth]{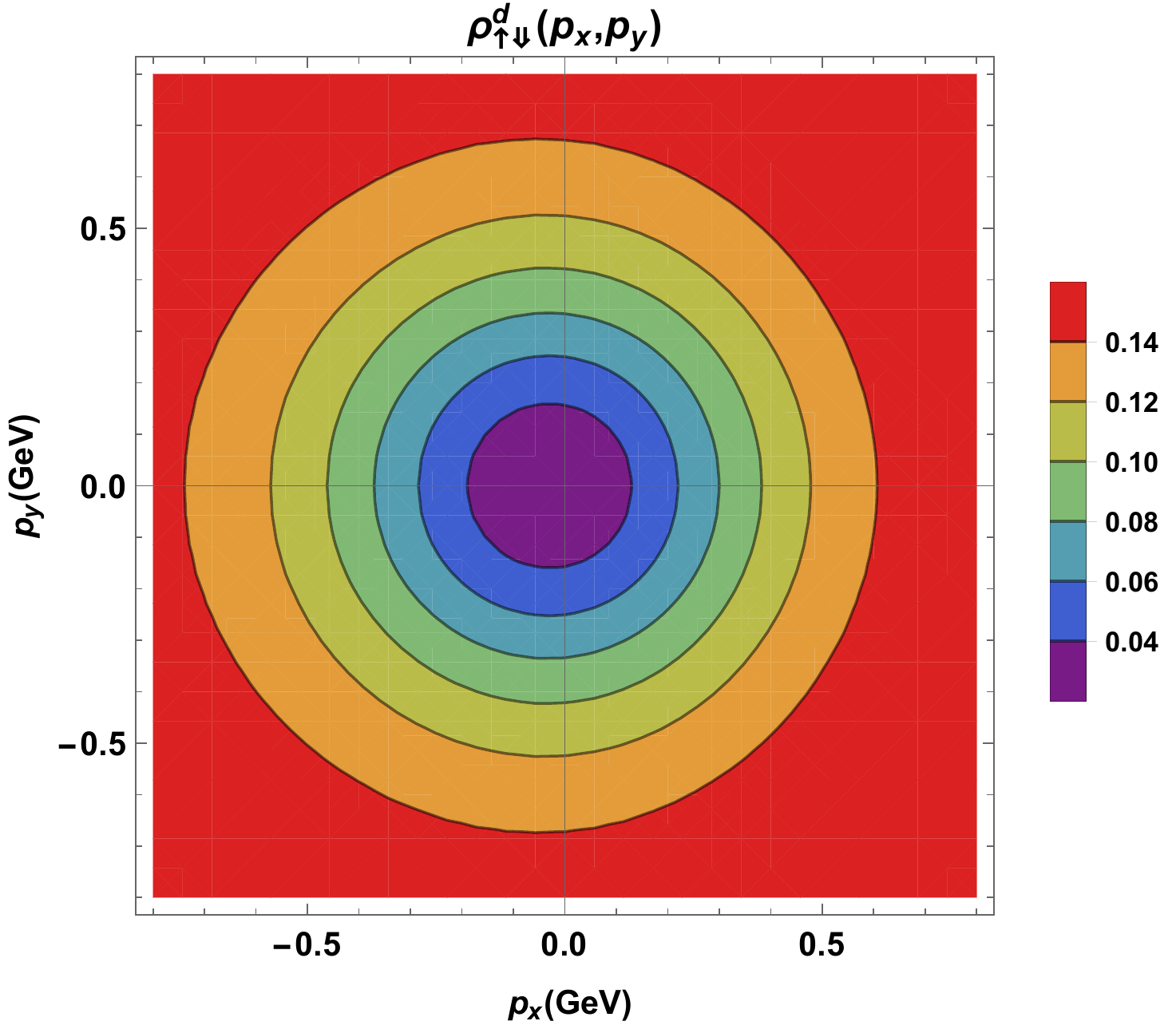}
\end{minipage}
\caption{The plot of Wigner distribution $\rho_{\uparrow \Downarrow}({\bf b}_\perp,{\bf p}_\perp)$ in transverse impact-parameter plane and transverse momentum plane for $u$-quark (left panel) and $d$-quark (right panel).}
\label{longi-up-trans-down}
\end{figure}
\begin{figure}
\centering
\begin{minipage}[c]{1\textwidth}
(a)\includegraphics[width=.4\textwidth]{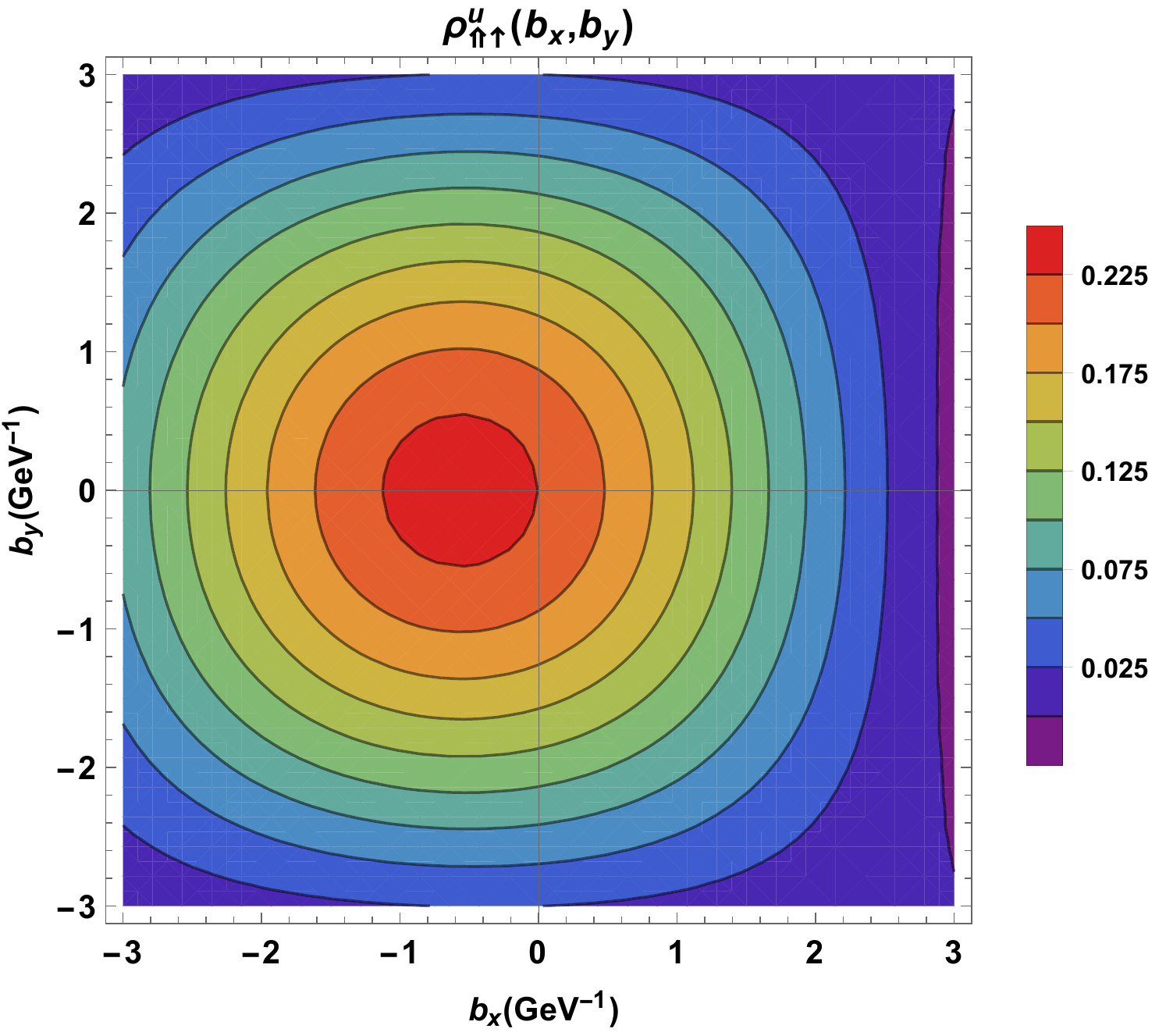}
(b)\includegraphics[width=.4\textwidth]{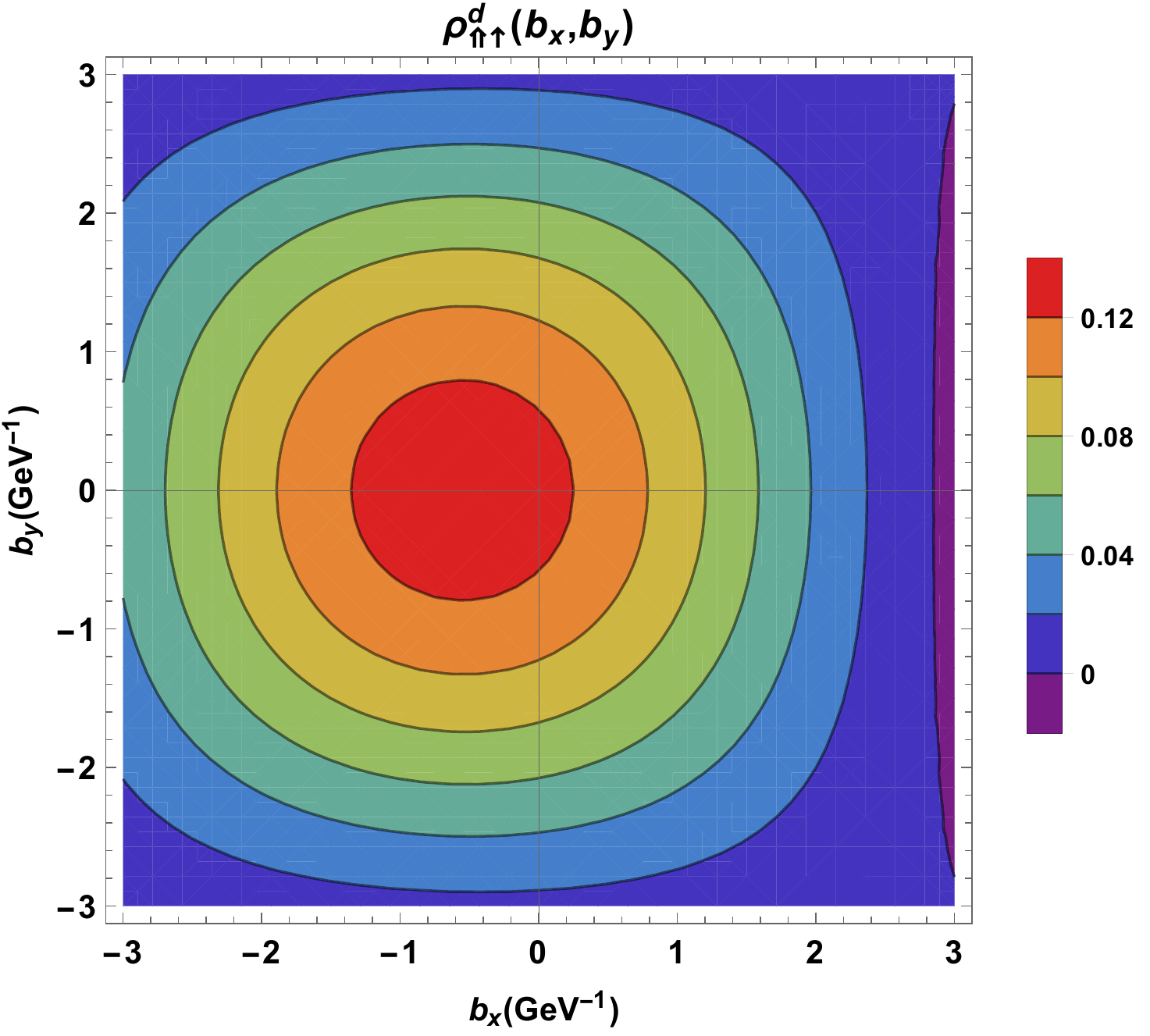}
\end{minipage}
\begin{minipage}[c]{1\textwidth}
(c)\includegraphics[width=.4\textwidth]{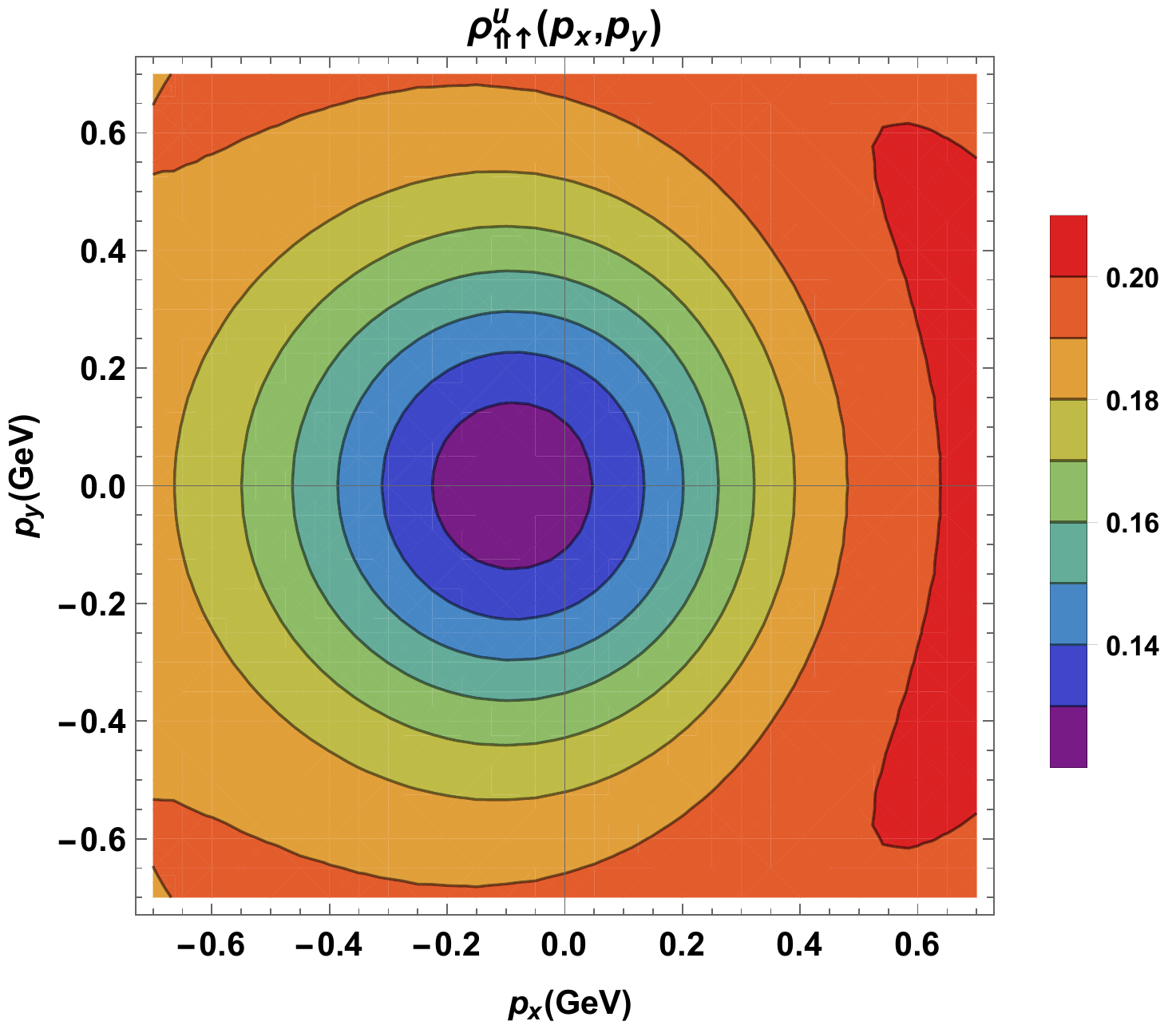}
(d)\includegraphics[width=.4\textwidth]{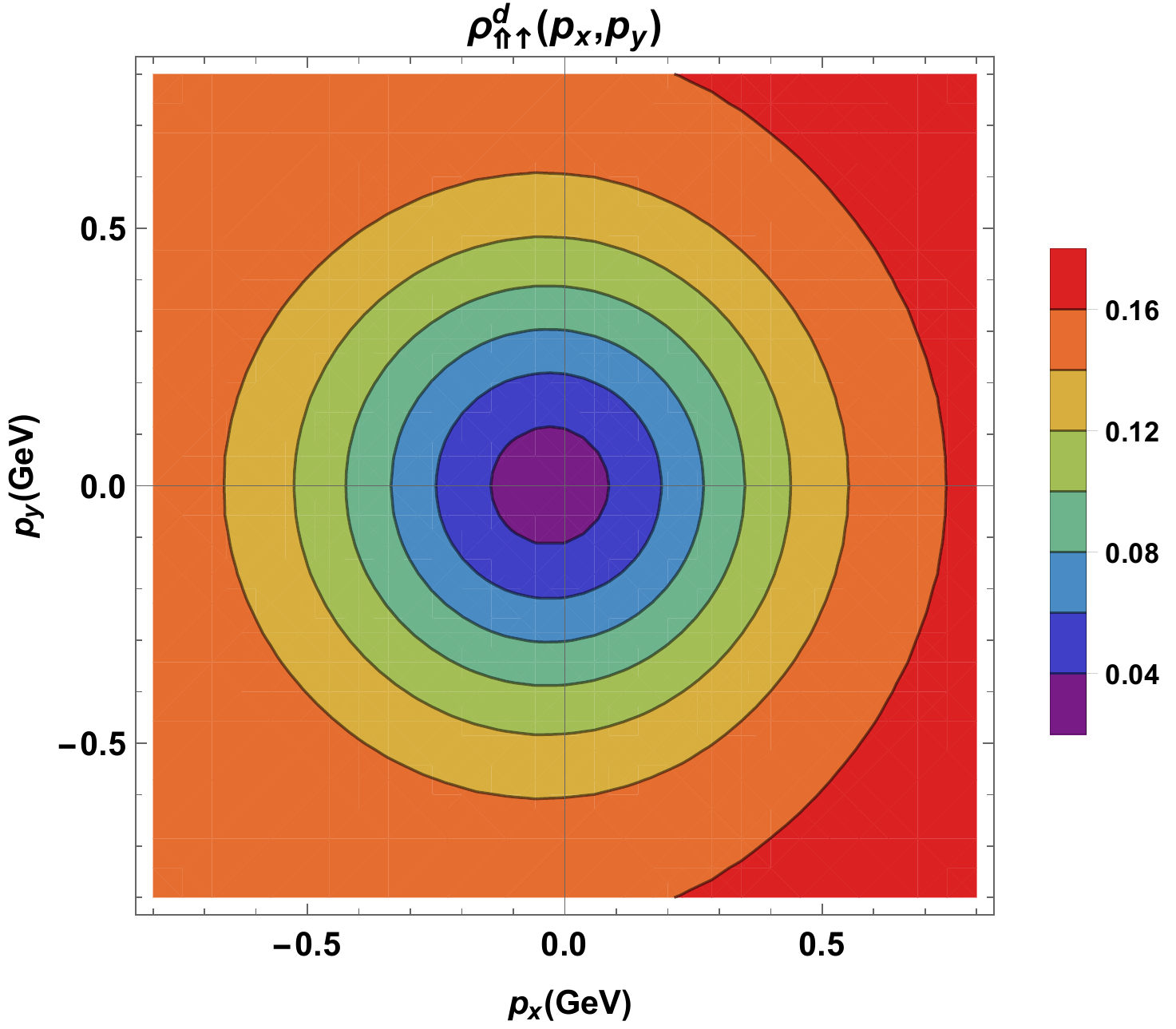}
\end{minipage}
\caption{The plot of Wigner distribution $\rho_{\Uparrow \uparrow}({\bf b}_\perp,{\bf p}_\perp)$ in transverse impact-parameter plane and transverse momentum plane for $u$-quark (left panel) and $d$-quark (right panel).}
\label{trans-up-longi-up}
\end{figure}
\begin{figure}
\centering
\begin{minipage}[c]{1\textwidth}
(a)\includegraphics[width=.4\textwidth]{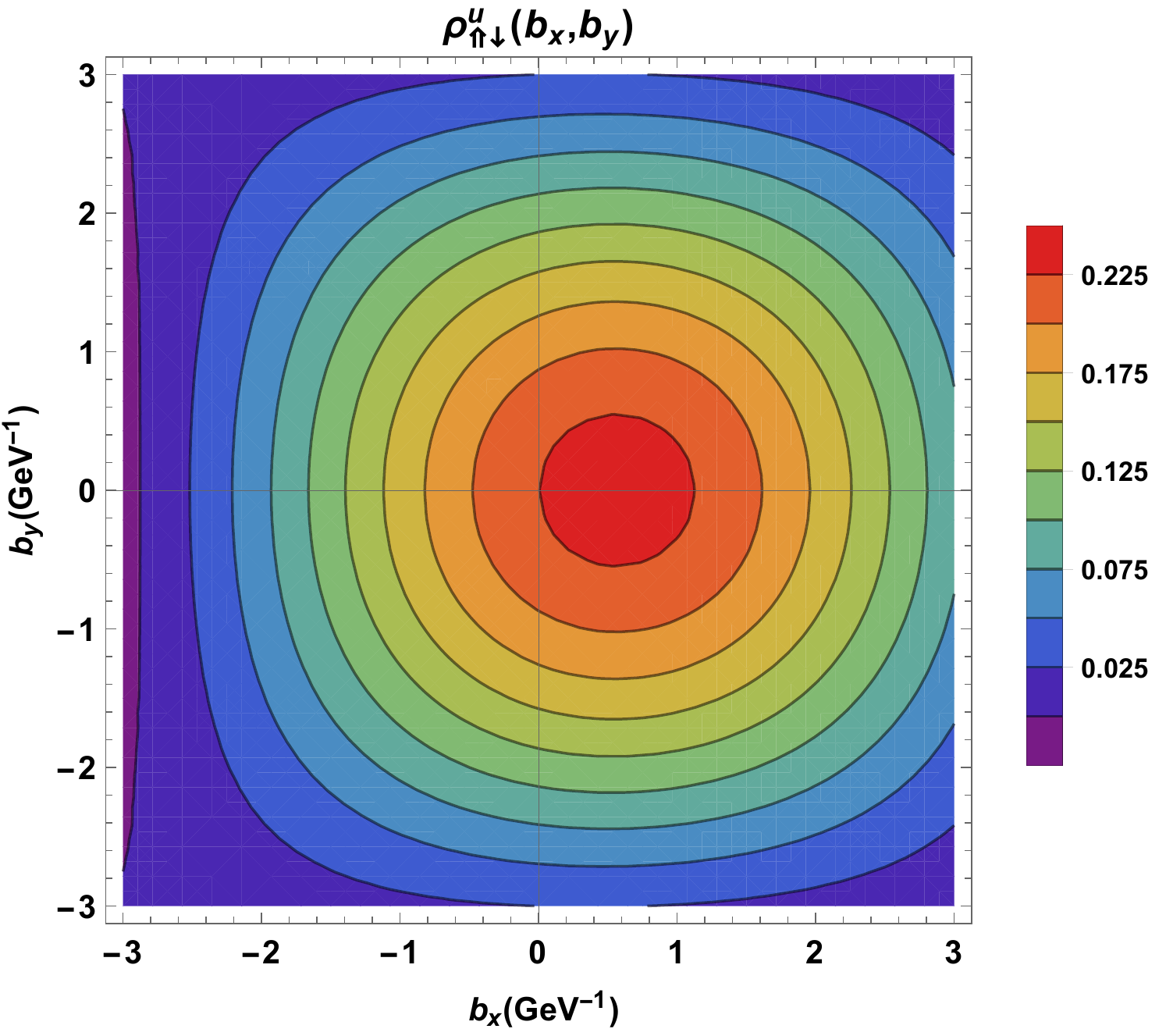}
(b)\includegraphics[width=.4\textwidth]{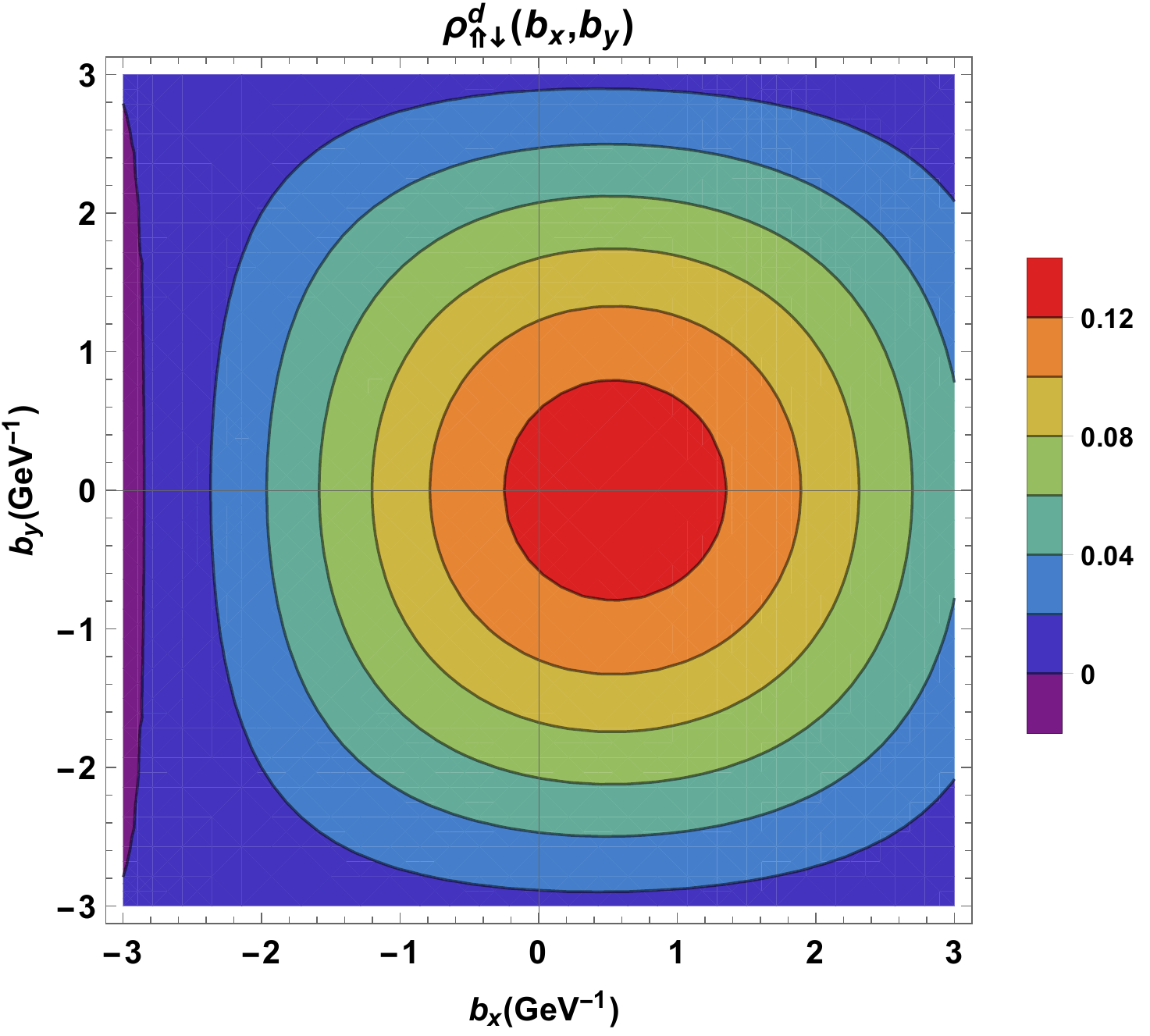}
\end{minipage}
\begin{minipage}[c]{1\textwidth}
(c)\includegraphics[width=.4\textwidth]{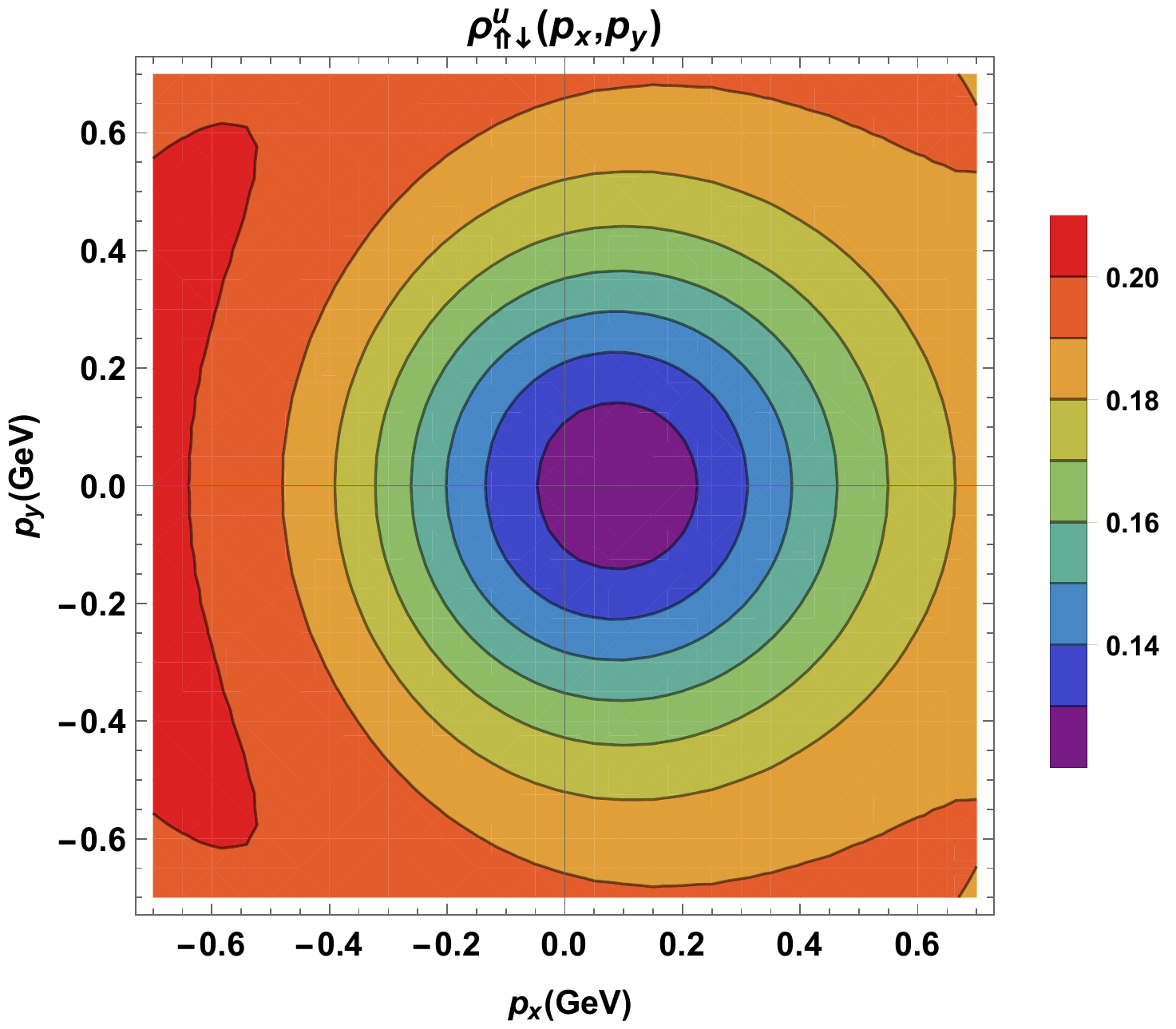}
(d)\includegraphics[width=.4\textwidth]{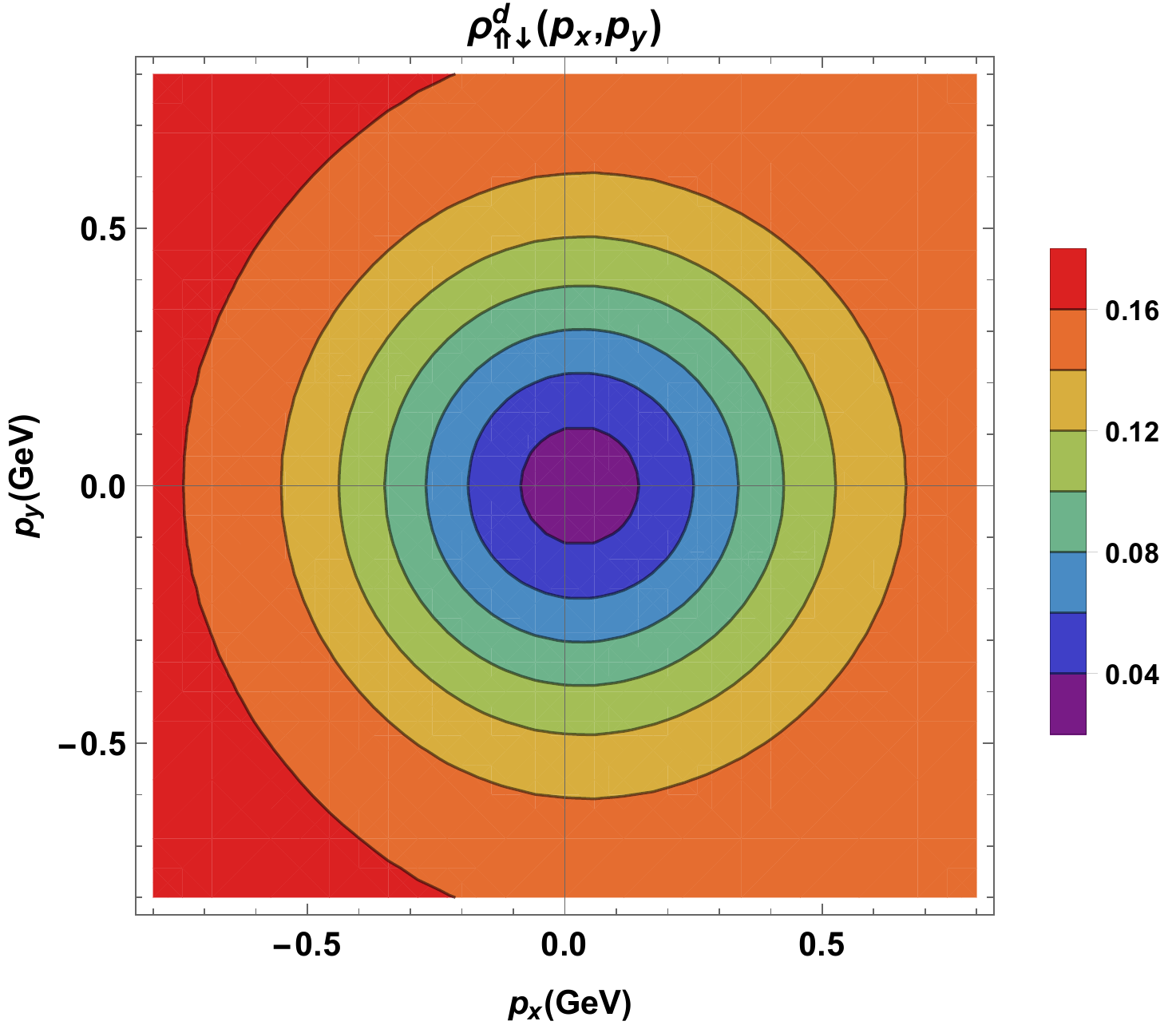}
\end{minipage}
\caption{The plot of Wigner distribution $\rho_{\Uparrow \downarrow}({\bf b}_\perp,{\bf p}_\perp)$ in transverse impact-parameter plane and transverse momentum plane for $u$-quark (left panel) and $d$-quark (right panel).}
\label{trans-up-longi-down}
\end{figure}

The transverse Wigner distribution have been plotted in Fig. \ref{trans-up-up}by for the case with  the quark having helicity as $\lambda=\Uparrow$ in the fermion composite system with helicity $\Lambda=\Uparrow$. The distortion in the distibution $\rho_{\Uparrow \Uparrow}$ shifts along the positive $b_y$ direction in impact-parameter plane for both $u$-quark and $d$-quark. In ${\bf p}_\perp$-plane, circularly symmteric distribution is observed, which is more focused at the center in case of $u$-quark, while it extends more to the higher values of transverse momentum of $d$-quark. For $\rho_{\Uparrow \Downarrow}$, the distortion is in opposite direction of $b_y$ for $u$-quark and $d$-quark when compared with $\rho_{\Uparrow \Uparrow}$ in impact-parameter plane (as shown in upper panels of Figs. \ref{trans-up-up} and \ref{trans-up-down}). In this work, we take the polarization direction of quark and proton along $x$-axis. From Eq. (\ref{transverse-wigner}) and Ref. \cite{wd8}, we find that in impact-parameter plane, the distortion comes due to $\rho^1_{UT}$ and $\rho^1_{TU}$, as they show dipolar distributions except the case of $\rho_{UU}$ and $\rho_{TT}$. However, in momentum plane, the strong correlation between the distributions $\rho_{UU}$, $\rho^1_{UT}$, $\rho^1_{TU}$ and $\rho_{TT}$, leads to the circular symmetric behaviour of distributions $\rho_{\Uparrow \Uparrow}$ and $\rho_{\Uparrow \Downarrow}$ for both quarks (as shown in lower panels of Figs. \ref{trans-up-up} and \ref{trans-up-down}).

Further, we plot the distribution $\rho_{\uparrow \Uparrow}$ in Fig. {\ref{longi-up-trans-up}}, which describes the correlation between spin of quark $\lambda_\perp=\Uparrow$ and spin of composite system $\lambda=\uparrow$. In impact-parameter plane, the distortion is clearly visible. This distortion is due to the Wigner distributions $\rho_{UT}$ and $\rho_{LU}$ as the dipolar distribution from these terms along $b_y$ and $b_x$ (shown in Ref. \cite{wd8}) adds up resulting  in $\rho_{\uparrow \Uparrow}$ in this model. Similarly, due to these terms, distortion is observed in ${\bf b}_\perp$-plane in case of $\rho_{\uparrow \Downarrow}$ as shown in Fig. \ref{longi-up-trans-down}. Because of the opposite transverse spin direction of quark in two cases, $\rho_{\uparrow \Downarrow}$ causes the distortion along negative $b_y$ while for $\rho_{\uparrow \Uparrow}$, it is in the direction of positive $b_y$. In momentum plane, we observe the distortion along negative $p_x$ for $u$-quark and $d$-quark when quark longitudinal spin direction is positive and proton transverse spin direction is positive (or negative). The observed distortion is more along negative $p_x$ in case of $d$-quark as compared to $u$-quark for $\rho_{\uparrow \Uparrow}$, while for $\rho_{\uparrow \Downarrow}$, it is more distorted in case of $u$-quark.

In Fig. \ref{trans-up-longi-up}, we plot the distribution $\rho_{\Lambda_\perp,\lambda}$, which explains the correlation between the transverse spin of composite system and longitudinal spin of quark, both along positive directions. Also the distortion comes from the correlation between the transverse spin of composite system along positive direction and longitudinal spin of quark along negative direction, shown in Fig. \ref{trans-up-longi-down}. From Eq. (\ref{trans-longi-distribution}), the distributions $\rho_{UU}$, $\rho^i_{TU}$, $\rho_{UL}$ and $\rho^i_{TL}$ are summed up according to the spin direction of composite system to get $\rho_{\Uparrow \uparrow}$ and $\rho_{\Uparrow, \downarrow}$. The resulting plotted distortion is along $b_x<0 (p_x<0)$ and $b_x>0(p_x>0)$ for $\rho_{\Uparrow \uparrow}$ and $\rho_{\Uparrow \downarrow}$ respectively for $u$ and $d$ quarks in ${\bf b}_\perp$-plane (${\bf p}_\perp$-plane).

\section{Conclusions}
We have presented the results of spin-spin correlations between the $u$-quark (or $d$-quark) and fermion composite system spins in light-front quark-diquark model evaluated from the Wigner distributions. The contribution from both the scalar and axial-vector diquarks is considered to get the distributions of $u$ and $d$ quarks. We consider the axial vector diquark to be further distinguished between the isoscalar or isovector depending upon the realistic analysis. First, we consider the spins of quark and fermion system in longitudinal direction i.e. $\lambda$ and $\Lambda$ respectively. Similarly, the correlation between transverse spin directions of quark $(\lambda_\perp)$ and composite particle $(\Lambda_\perp)$ is evaluated. Further, the different combinations are taken into account i.e. when quark spin is in longitudinal direction and spin of composite particle is in transverse direction and vice-versa i.e. $\rho_{\Lambda_\perp \lambda}$ and $\rho_{\Lambda \lambda_\perp}$. All these results are presented in transverse impact-parameter plane (${\bf b}_\perp$-plane) and transverse momentum plane (${\bf p}_\perp$-plane). We observe that the distortions in the correlations seen in both planes are due to the effect of different Wigner distributions.

The spin-spin correlations are related to the Wigner distributions and the quantum mechanical version of quark Wigner distributions have not yet been measured experimentally. The measurable quantities can be extracted from Wigner distributions by integrating them over transverse position or transverse momentum of quark. These quantities can be experimentally measured via DVCS or Drell-Yan processes. Further, since the Wigner distributions are related to  GTMDs through Fourier transformations, they can be accessible through exclusive double Drell-Yan process.
   
\acknowledgements
H.D. would like to thank the Department of Science and Technology (Ref No.
EMR/2017/001549) Government of India for financial support.

\end{document}